\documentclass[a4paper,traditabstract]{aa}
\usepackage{amsmath}
\usepackage{graphicx}
\usepackage{txfonts}
\usepackage{natbib}
\usepackage{subfig}
\usepackage{color}
\usepackage{mathrsfs}
\usepackage{xspace}

\usepackage[plainpages=False,
            bookmarks=False,
            colorlinks=True,
            citecolor=blue,
            linkcolor=blue]{hyperref}

\graphicspath{{Figures/}}

\newcommand{\scname}[1]{\textsc{#1}}

\newcommand{\snifs}{\scname{SNIFS}\xspace}
\newcommand{\snf}{\scname{SNfactory}\xspace}

\newcommand{\skp}{\emph{SkyProbe}\xspace}
\newcommand{\aeronet}{\scname{AERONET}\xspace}

\newcommand{\sn}{supernova\xspace}
\newcommand{\sne}{supernov\ae\xspace}

\newcommand{\sneia}{{SNe~Ia}\xspace}

\newcommand{\cf}{\emph{cf.}\xspace}      % confere
\newcommand{\eg}{\emph{e.g.}\xspace}     % exempli gratia
\newcommand{\ie}{\emph{i.e.}\xspace}     % id est
\newcommand{\HdO}{\ensuremath{\mathrm{H_{2}O}}\xspace}
\newcommand{\Od}{\ensuremath{\mathrm{O_{2}}}\xspace}
\newcommand{\Of}{\ensuremath{\mathrm{O_{4}}}\xspace}
\newcommand{\Oz}{\ensuremath{\mathrm{O_{3}}}\xspace}

\newcommand{\atm}{\ensuremath{\mathrm{atm}}} % Atmosphere
\newcommand{\tel}{\ensuremath{\mathrm{tel}}} % Telescope
\newcommand{\tell}{\ensuremath{\oplus}}      % Telluric
\newcommand{\ang}{\AA{}ngstr\"om\xspace}

\begin{document}

% Title
\title {Atmospheric extinction properties above Mauna Kea from the
  Nearby Supernova Factory spectro-photometric data set}

\titlerunning{Mauna Kea atmospheric extinction properties}
\authorrunning{C.~Buton \& the \snf}

% Authors
\author{
  The Nearby Supernova Factory: \\
  C.~Buton\inst{1},         % PI Bonn
  Y.~Copin\inst{2},         % IPNL
  G.~Aldering\inst{3},      % LBNL
  P.~Antilogus\inst{4},     % LPNHE
  C.~Aragon\inst{3},        % LBNL
  S.~Bailey\inst{3},        % LBNL
  C.~Baltay\inst{5},        % Yale University
  S.~Bongard\inst{4},       % LPNHE
  A.~Canto\inst{4},         % LPNHE
  F.~Cellier-Holzem\inst{4}, % LPNHE
  M.~Childress\inst{3,6},   % LBNL, UC Berkeley
  N.~Chotard\inst{7,8},     % Tsinghua & NAOC
  H.~K.~Fakhouri\inst{3,6}, % LBNL, UC Berkeley
  E.~Gangler\inst{2},       % IPNL
  J.~Guy\inst{4},           % LPNHE
  E.~Y.~Hsiao\inst{3},      % LBNL
  M.~Kerschhaggl\inst{1},   % PI Bonn
  M.~Kowalski\inst{1},      % PI Bonn
  S.~Loken\inst{3}\thanks{Deceased.}, % LBNL
  P.~Nugent\inst{9,10},     % Computational Center, UC astro
  K.~Paech\inst{1},         % PI Bonn
  R.~Pain\inst{4},          % LPNHE
  E.~P\'econtal\inst{11},   % CRAL
  R.~Pereira\inst{2},       % IPNL
  S.~Perlmutter\inst{3,6},  % LBNL, UC Berkeley
  D.~Rabinowitz\inst{5},    % Yale University
  M.~Rigault\inst{2},       % IPNL
  K.~Runge\inst{3},         % LBNL
  R.~Scalzo\inst{5,12},     % Yale University, ANU
  G.~Smadja\inst{2},        % IPNL
  C.~Tao\inst{13,7},        % CPPM, Tsinghua
  R.~C.~Thomas\inst{9},     % Computational Center
  B.~A.~Weaver\inst{14},    % NYU
  and C.~Wu\inst{4,8}       % LPNHE, NAOC
}

\institute{%
  Physikalisches Institut Universit\"at Bonn (Bonn), Nussallee 12, 53115
  Bonn, Germany
  \and
  Universit\'e de Lyon, F-69622, Lyon, France; Universit\'e Lyon 1,
  Villeurbanne; CNRS/IN2P3, Institut de Physique Nucl\'eaire de Lyon.
  \and
  Physics Division, Lawrence Berkeley National Laboratory, 1 Cyclotron
  Road, Berkeley, CA, 94720
  \and
  Laboratoire de Physique Nucl\'eaire  et des Hautes \'Energies,
  Universit\'e Pierre et Marie Curie Paris 6, Universit\'e Paris Diderot
  Paris 7, CNRS-IN2P3, 4 place Jussieu, 75252 Paris Cedex 05, France
  \and
  Department of Physics, Yale University, New Haven, CT 06250-8121
  \and
  Department of Physics, University of California Berkeley, 366 LeConte
  Hall MC 7300, Berkeley, CA, 94720-7300
  \and
  Tsinghua Center for Astrophysics, Tsinghua University, Beijing 100084,
  China
  \and
  National Astronomical Observatories, Chinese Academy of Sciences,
  Beijing 100012, China
  \and
  Computational Cosmology Center, Computational Research Division,
  Lawrence Berkeley National Laboratory, 1 Cyclotron Road MS 50B-4206,
  Berkeley, CA, 94720
  \and
  Department of Astronomy, University of California, Berkeley, CA
  94720-3411, USA
  \and
  Centre de Recherche Astronomique de Lyon, Universit\'e Lyon 1, 9
  Avenue Charles Andr\'e, 69561 Saint Genis Laval Cedex, France
  \and
  Research School of Astronomy \& Astrophysics, The Australian National
  University, Mt.  Stromlo Observatory, Cotter Road, Weston Creek, ACT
  2611 Australia
  \and
  CPPM, 163 Av.  Luminy, 13288 Marseille Cedex 09, France
  \and
  New York University, Center for Cosmology and Particle Physics, 4
  Washington Place, New York, NY 10003}

% Dates
\date{Accepted for publication (A\&A) on October 2nd, 2012}

% Abstract
\abstract{We present  a new atmospheric  extinction curve for  Mauna Kea
  spanning 3200--9700~\AA.  It is the most comprehensive  to date, being
  based on some 4285~standard star spectra obtained on 478~nights spread
  over  a period  of 7~years  obtained by  the Nearby  SuperNova Factory
  using the  SuperNova Integral Field Spectrograph. This  mean curve and
  its dispersion can  be used as an aid  in calibrating spectroscopic or
  imaging  data  from  Mauna  Kea,  and in  estimating  the  calibration
  uncertainty associated  with the use  of a mean extinction  curve. Our
  method for decomposing the  extinction curve into physical components,
  and the ability  to determine the chromatic portion  of the extinction
  even on cloudy  nights, is described and verified  over the wide range
  of  conditions  sampled by  our  large  dataset.  We demonstrate  good
  agreement  with atmospheric science  data obtain  at nearby  Mauna Loa
  Observatory,  and  with   previously  published  measurements  of  the
  extinction above Mauna Kea.}

% Keywords

\maketitle

%%%%%%%%%%%%%%%%%%%%%%%%%%%%%%%%%%%%%%%%%%%%%%%%%%%%%%%%%%%%%%%%%%%%%%

\section{Introduction}
\label{sec:introduction}

The summit of Mauna Kea in Hawaii is home to the largest and most
powerful collection of astronomical telescopes in the world. For many
studies accurate flux calibration is critical for deriving the maximum
amount of information from observations with these telescopes, and
correction for the optical atmospheric and instrumental transmissions is
one of the main limitations of astronomical flux measurements from the
ground \citep[see][for the most recent reviews]{Burke10,
  Patat11}. Therefore, as part of our Nearby SuperNova Factory project
\citep[\snf,][]{Aldering02} we have carefully monitored the atmospheric
transmission over the course of our observing campaign, and in this
paper describe findings that should be of use to other Mauna Kea
observers.

According to the GONG (Global Oscillation Network Group) site survey
\citep{Hill94a, Hill94b}, the extinction above the summit of Mauna Kea
is among the lowest and most stable of any astronomical site. Several
studies were carried out at the end of the 80's to assess the
atmospheric characteristics above Mauna Kea \citep{Krisciunas87,
  Boulade87, Beland88}. These have formed the basis for the standard
Mauna Kea extinction curve provided by most observatories on Mauna
Kea. However, the results reported by \cite{Boulade87} and
\cite{Beland88} are single-night extinction studies carried out using
the 3.6~m Canada France Hawaii Telescope (CFHT) over limited wavelength
ranges (3100--3900~\AA{} and 3650--5850~\AA, respectively). Thus, they
do not cover the entire optical window, nor do they reflect variability
of the extinction. The measurements of the optical extinction from the
Mauna Kea summit presented in \cite{Krisciunas87} are based on 27~nights
of $B$ and $V$ band ($\sim 4400$~\AA{} and $\sim 5500$~\AA{})
measurements from three different telescopes, including the 2.2~m
University of Hawaii telescope (UH88) and the CFHT, between 1983 and
1985.  Since then, only the evaluation of the quality of the site for
the Thirty Meter Telescope (TMT) has been published \citep{Schock09,
  Travouillon11}. The TMT site testing campaign confirms that Mauna Kea
is one of the best sites for ground based astronomy but does not include
the properties and the variability of the spectral extinction at the
site.

\snf \citep{Aldering02} was developed for the study of dark energy using
Type Ia supernovae (\sneia).  Our goal has been to find and study a
large sample of nearby \sneia, and to achieve the percent-level
spectro-photometric calibration necessary so that these \sneia can be
compared with \sneia at high redshifts.  Since 2004 the \snf has
obtained spectral time series of over 200 thermonuclear \sne and these
are being used to measure the cosmological parameters and improve \sneia
standardization by empirical means and through a better understanding of
the underlying physics.  The main asset of the \snf collaboration is the
SuperNova Integral Field Spectrograph \citep[\snifs,][]{Lantz04}, a
dedicated integral field spectrograph built by the collaboration and
mounted on the University of Hawaii UH88 telescope.

Along with the \sn observations, the \snf data set includes
spectro-photometric observations of standard stars with \snifs, which
are used to obtain the instrumental calibration and the atmospheric
extinction.  The observations benefit from a large wavelength range
(3200--9700~\AA) and the high relative precision that \snifs can
achieve. When possible, standards were obtained throughout a given night
to help distinguish between spectral and temporal variations of the
transmission. While it is common practice to derive an extinction curve
by solving independently for the extinction at each wavelength, this
approach ignores the known physical properties of the atmosphere, which
are correlated across wide wavelength regions.  Using the standard star
spectra to disentangle the physical components of the atmosphere
extinction ensures a physically meaningful result, allows for robust
interpolation across standard star spectral features, and provides a
simpler and more robust means of estimating the error covariance matrix.
The method described in this paper allows us to obtain such a complete
atmospheric model for each observation night, {\it including nights
  afflicted by clouds}. We use the results to generate a new mean Mauna
Kea extinction curve, and to explore issues related to variation in the
extinction above Mauna Kea.

\section{The \snf Mauna Kea extinction dataset}
\label{sec:extinction_dataset}

We begin by describing the basic properties of the dataset to be used in
measuring the extinction properties above Mauna Kea.

\subsection{The Supernova Integral Field Spectrograph \& data
  reduction}
\label{sec:snifs}

\snifs is a fully integrated instrument optimized for automated
observation of point sources on a structured background over the full
optical window at moderate spectral resolution.  It consists of a
high-throughput wide-band pure-lenslet integral field spectrograph
\citep[IFS, ``\`a la TIGER'';][]{Bacon95, bacon01}, a multifilter
photometric channel to image the field surrounding the IFS for
atmospheric transmission monitoring simultaneous with spectroscopy, and
an acquisition/guiding channel.  The IFS possesses a fully filled
$6\farcs 4 \times 6\farcs 4$ spectroscopic field of view subdivided into
a grid of $15 \times 15$ spatial elements (spaxels), a dual-channel
spectrograph covering 3200--5200~\AA{} and 5100--9700~\AA{}
simultaneously, and an internal calibration unit (continuum and arc
lamps).  \snifs is continuously mounted on the south bent Cassegrain
port of the University of Hawaii 2.2~m telescope on Mauna Kea, and is
operated remotely.  The \snifs standard star spectra were reduced using
our dedicated data reduction procedure, similar to that presented in
Section~4 of \cite{bacon01}.  A brief discussion of the spectrographic
pipeline was presented in \cite{Aldering06}. Here we outline changes to
the pipeline since that work, but leave a complete discussion of the
reduction pipeline to subsequent publications focused on the instrument
itself.

After standard CCD preprocessing and subtraction of a low-amplitude
scattered-light component, the 225~spectra from the individual spaxels
of each SNIFS exposure are extracted from each blue and red spectrograph
exposure, and re-packed into two $(x,y,\lambda)$-datacubes.  This highly
specific extraction is based upon a detailed optical model of the
instrument including interspectrum crosstalk corrections.  The datacubes
are then wavelength-calibrated using arc lamp exposures acquired
immediately after the science exposures, and spectro-spatially
flat-fielded using continuum lamp exposures obtained during the same
night. Cosmic rays are detected and corrected using a
three-dimensional-filtering scheme applied to the datacubes.

Standard star spectra are extracted from each $(x,y,\lambda)$-datacube
using a chromatic spatial point-spread function (PSF) fit over a uniform
background
\citep{Buton09}\footnote{\url{http://tel.archives-ouvertes.fr/docs/00/56/62/31/PDF/TH2009_Buton_ClA_ment.pdf}}.
The PSF is modeled semi-analytically as a constrained sum of a Gaussian
(describing the core) and a Moffat function (simulating the wings).  The
correlations between the different shape parameters, as well as their
wavelength dependence, were trained on a set of 300~standard star
observations in various conditions of seeing and telescope focus between
2004 and 2007 with \snifs.  This resulted in a empirical chromatic model
of the PSF, depending only on an effective width (accounting for seeing)
and a flattening parameter (accounting for small imaging defocus and
guiding errors). The PSF modeling properly takes the
wavelength-dependent position shift induced by atmospheric differential
refraction into account without resampling.

\subsection{Data characteristics and sub-sample selection}
\label{sec:dataset}

\begin{figure}
  \centering
  \subfloat[Number of standard stars per night]{%
    \label{fig:standards-number}%
    \includegraphics[width=\columnwidth]{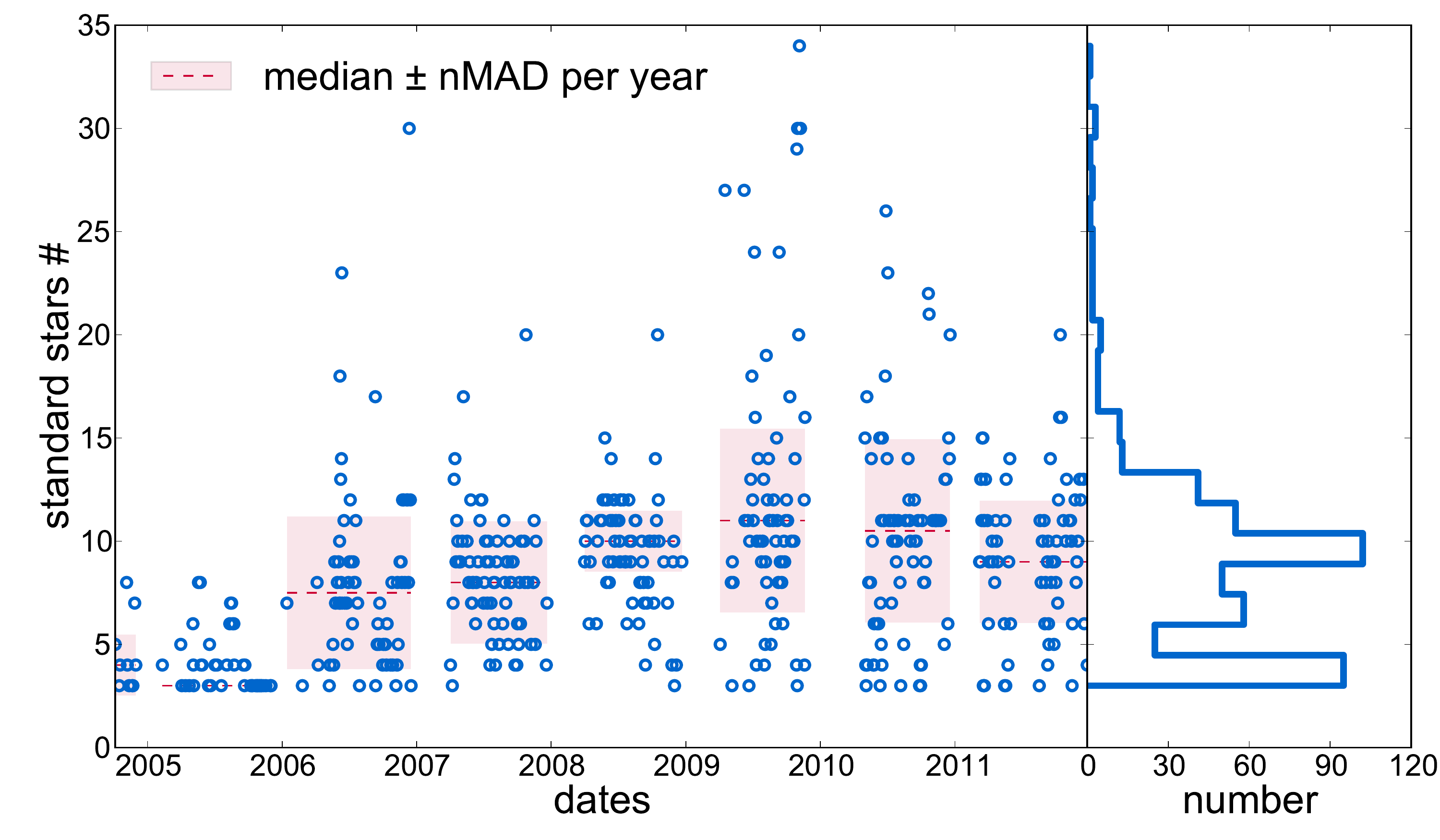}}%
  \hspace{0mm}%
  \subfloat[Airmass range ($\Delta_{X}$) per night]{%
    \label{fig:standards-airmass}%
    \includegraphics[width=\columnwidth]{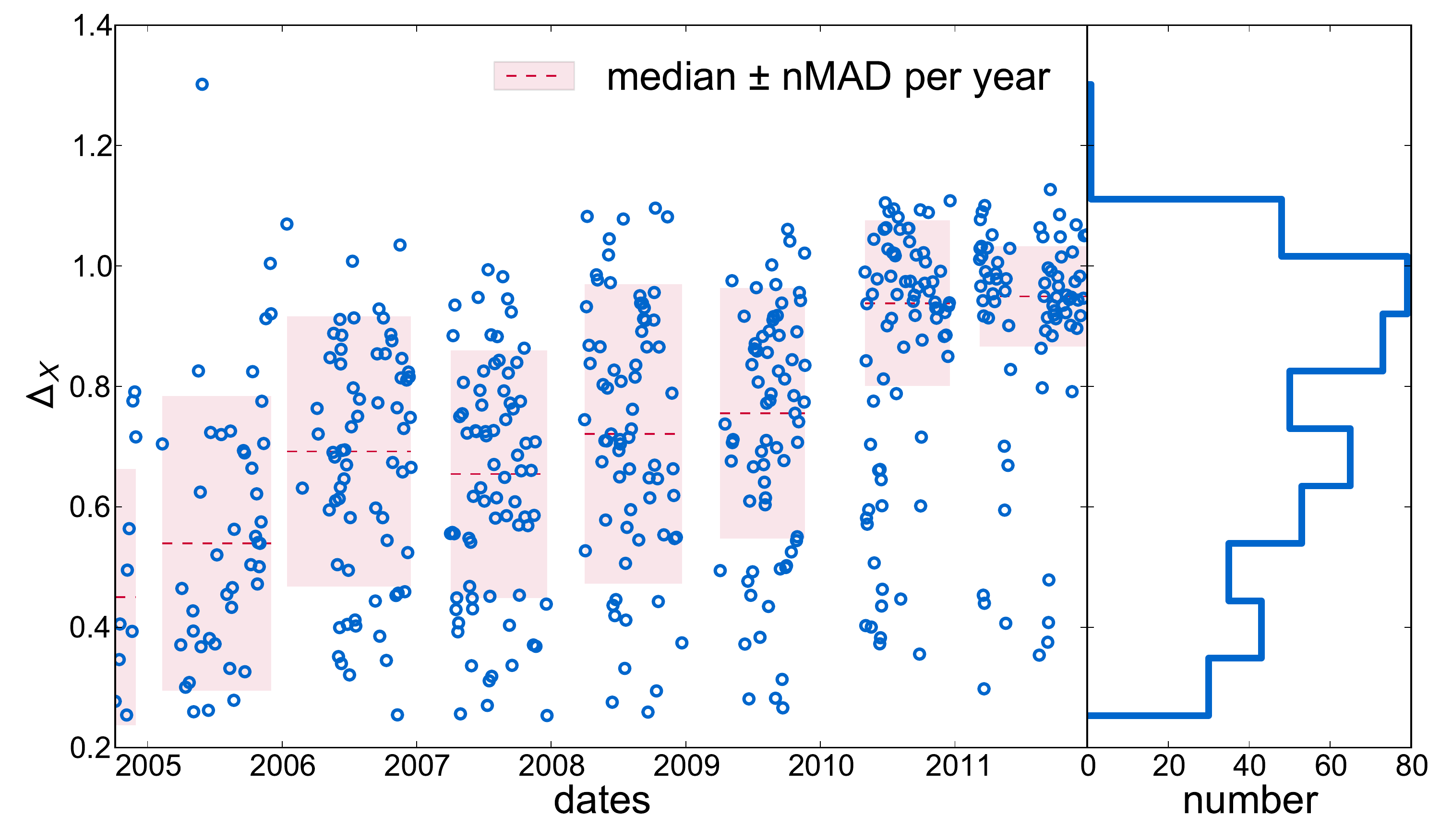}}%
  \caption{Time evolution of the standard star number (\# standards) per
    night \protect\subref{fig:standards-number} and of the airmass range
    ($\Delta_{X}$) per night \protect\subref{fig:standards-airmass}
    after quality cuts. The nMAD acronym stands for ``normalized median
    absolute deviation'' (the normalization factor is introduced in
    order to use the median absolute deviation as a consistent estimator
    for the estimation of the standard deviation).}
  \label{fig:standards-statistics}
\end{figure}

The \snf spectro-photometric follow-up has been running regularly since
September 2004, with a periodicity of two to three nights. In most years
observations were concentrated in the April to December period in order
to coincide with the best weather at Palomar where we carried out our
search for \sne. Initially the nights were split with UH observers, with
\snf taking the second half of allocated nights. In May 2006, our
program switched from half-night to full-night observations.  The \snf
time was mainly used to observe \sneia (up to 20 per night), but in
order to flux calibrate the \sn data, standard star observations were
inserted throughout the night.

Two different kinds of standard stars were observed: bright standard
stars ($V = 4\text{--}7$) were mainly observed during nautical twilight,
while faint standard stars ($V = 10\text{--}14$) were observed during
astronomical twilight and during the night.  A list of the standard
stars used for calibration purposes by the \snf is given in
Table~\ref{tab:standard-stars}.  A typical night started with $\sim 3$
bright standard stars and one faint standard star during evening
twilight, followed by 3 to 4 faint standard stars distributed all along
the night in between \sne, and finished with another faint standard and
$\sim 3$ bright standard stars during morning twilight. Generally the
calibration provided in the literature for the fainter stars is of
higher quality. Moreover, we found that the very short (1--2 second)
exposures required for bright standard stars resulted in very complex
PSF shapes. During the period when observations were conducted only
during the second half of the night, the typical number of standard
stars was more limited, as seen in Fig.~\ref{fig:standards-number}.

The evolution through the years in the number of standard stars observed
per night is shown in Fig.~\ref{fig:standards-number}, while
Fig.~\ref{fig:standards-airmass} shows the airmass range. The noticeable
changes in the numbers and airmass distribution have several causes. As
mentioned above, the half-night allocations up through May 2006
restricted the number of standard stars that could be observed each
night without adversely impacting our SN program. In addition, in order
to improve the automation in executing our observing program we
developed algorithms to select a standard star during windows slated for
standard star observations. Initially the selections were pre-planned
manually each night, but this approach was not sufficiently flexible,
\eg to account for weather interruptions. In fall 2005 we developed an
automated selection algorithm designed to pick an observable star that
was best able to decorrelate airmass and spectral type based on the
standard stars previously observed that night. The idea here was to
obtain a good airmass range and observe enough different stellar
spectral classes to avoid propagating stellar features into the
calibration. More recently, having convinced ourselves that stellar
features did not present a problem when using the physical extinction
model presented here, the automated selection was changed so as to
minimize the extinction uncertainty, considering the standard stars
previously observed that night. Occasionally, nights when the Moon was
full were dedicated to intensive observations of standard stars as a
means of testing various aspects of our method and software
implementation. Finally, a few inappropriate standard stars --
Hiltner600 (double star), HZ4, LTT7987, GRW+705824 (broad-line DA white
dwarfs) -- have been phased out and are not included in the analysis
presented here.

Of the 711~nights originally available with spectroscopic data, we
removed 77~nights with only one or no standard star available. These
occurred primarily due to unplanned telescope closures during the night
caused by weather or technical problems. Such cases were concentrated
during the period when only half-nights were available.  For these
nights, since an extinction solution is not possible, a mean atmospheric
extinction is used for the flux calibration of the science targets. To
ensure the quality of the Mauna Kea extinction determination for the
present study, we also choose to discard nights with an \emph{expected}
extinction error larger than 0.1~magnitude/airmass; this resulted in the
exclusion of an additional 55~nights. The expected extinction accuracy
was calculated using the known airmass distribution of the standard
stars and using the achromatic extraction error of 3\% and 2\%
empirically found for bright and faint standard stars, respectively
\citep{Buton09}.  Calibration of science data on such nights is not
necessarily a problem, it is simply that the atmospheric properties are
difficult to decouple from the instrument calibration on such
nights. Finally, strict quality cuts on the number of stars ($\geq 3$)
and airmass range ($\geq 0.25$) per night are applied (respectively 68
and 33~nights are skipped). Additional cuts based on flags from the
pre-processing steps and the quality of the fit were applied to avoid
bad exposures or spectra with production issues. In the end, the data
sample is comprised of 4285~spectra from 478~nights that passed these
very restrictive cuts.

\section{Flux calibration formalism}
\label{sec:formalism}

In a given night the spectrum, $S_{i}(\lambda, \hat{z}_{i},
t)$\footnote{expressed in pseudo-ADU/s/\AA}, of an astronomical source
$i$ observed by \snifs can be expressed as,
\begin{equation}
  \label{eq:formalism-1}
  S_{i}(\lambda, \hat{z}_{i}, t) = S^{\star}_{i}(\lambda, t) \times
  C(\lambda, t) \times T_{\atm}(\lambda, \hat{z}_{i}, t),
\end{equation}
where $S^{\star}_{i}(\lambda, t)$ is the intrinsic spectrum of the
source\footnote{expressed in erg/cm$^{2}$/s/\AA} as it would be seen
from above the atmosphere, $T_{\atm}(\lambda, \hat{z}, t)$ is the
time-dependent, line-of-sight ($\hat{z}$) dependent atmospheric
transmission.  $C(\lambda, t)$ is the instrument calibration (\ie the
combined response of the telescope, the instrument and the detector),
such that,
\begin{equation}
  \label{eq:formalism-2}
  C(\lambda, t) = T_{\tel}(\lambda, t) \times T_{\snifs}(\lambda, t) \times
  Q(\lambda, t)
\end{equation}
where $T_{\tel}(\lambda, t)$, $T_{\snifs}(\lambda, t)$ and $Q(\lambda,
t)$ are respectively the chromatic telescope transmission, instrument
transmission and detector quantum efficiency, all of which are
potentially time dependent.

Because the data are divided by a flat-field exposure that is not
required to be the same from night to night, we are interested only in
$t$ spanning one night intervals. We will assume that the instrument
response is stable over the course of a night, and therefore write:
\begin{equation}
  \label{eq:formalism-3}
  C(\lambda, t) = C(\lambda).
\end{equation}
Later, in \S~\ref{sec:stability-instrument}, we re-examine this question
and confirm that it is valid at a level much better than 1\%.

As for the atmospheric extinction, we choose to separate
$T_{\atm}(\lambda, \hat{z}_{i}, t)$ with respect to its time dependence,
as follows:
\begin{equation}
  \label{eq:formalism-4}
  T_{\atm}(\lambda, \hat{z}_{i}, t) =
  \overline{T}_{\atm}(\lambda, \hat{z}_{i}) \times
  \delta T_{i}(\lambda, \hat{z}_{i}, t).
\end{equation}
Here $\delta T_{i}(\lambda, \hat{z_{i}}, t)$ represents the normalized
atmospheric transmission variability at the time, $t$, along the line of
sight, $\hat{z}_{i}$, to the star $i$. By definition, a photometric
night is one in which the $\delta T_{i}$ are retrospectively found to be
compatible with 1 (taking into account the measurement errors),
irrespectively of wavelength, direction, or time.

Furthermore, it is common in astronomy to express the extinction in
magnitudes, such that the transmission, $\overline{T}_{\atm}(\lambda,
\hat{z})$, is given by
\begin{equation}
  \label{eq:formalism-6}
  \overline{T}_{\atm}(\lambda, \hat{z}) =
  10^{-0.4 \times K_{\atm}(\lambda, \hat{z})},
\end{equation}
where $K_{\atm}(\lambda, \hat{z})$ is the atmospheric extinction in
magnitudes per airmass.

Overall, Eq.~\eqref{eq:formalism-1} becomes:
\begin{equation}
  \label{eq:formalism-7}
  \log \frac{S_{i}(\lambda, \hat{z}_{i}, t)}{S^{\star}_{i}(\lambda, t)} =
  \log C(\lambda) - 0.4 \times K_{\atm}(\lambda, \hat{z}_{i}) +
  \log \delta T_{i}(\lambda, \hat{z}_{i}, t).
\end{equation}

For standard star observations, $S^{\star}(\lambda, t)$ is supposedly
known (\cf \S~\ref{sec:results}) and the unknowns are $C(\lambda)$,
$K_{\atm}(\lambda, \hat{z})$ and the $\delta T_{i}(\lambda, \hat{z}_{i},
t)$ (one per star $i$). Conversely, for a supernova observation,
$S^{\star}(\lambda, t)$ becomes the unknown, and $C(\lambda)$, $
K_{\atm}(\lambda, \hat{z})$ and any deviation of $\delta T_{i}(\lambda,
\hat{z}, t)$ from unity would need to be known in order to achieve flux
calibration. As outlined in \cite{Aldering02} and \cite{pereira08}, with
SNIFS, $\delta T_{i}(\lambda, \hat{z}, t)$ can be determined from
secondary stars on the parallel imaging channel for fields having at
least one visit on a photometric night.

Our focus in this paper is on the properties of the atmospheric
extinction, $K_{\atm}(\lambda, \hat{z})$. But as we have just seen, its
determination is linked to the determination of the instrument
calibration, $C(\lambda)$, and of any atmospheric transmission
variations with time, $\delta T_{i}(\lambda, \hat{z}_{i}, t)$, for each
standard star, $i$. In order to constrain the extinction,
$K_{\atm}(\lambda, \hat{z})$, to have a meaningful shape, we now present
its decomposition into physical components.

\section{Atmospheric extinction model}
\label{sec:atmospheric-extinction}

It is now well established that the wavelength dependence of the
atmospheric extinction, $K_{\atm}(\lambda, \hat{z})$, is the sum of
physical elementary components \citep{Hayes75, Wade88, Stubbs07}, either
scattering or absorption.  Furthermore, the extinction increases with
respect to airmass $X$ along the line of sight, $\hat{z}$, giving:
\begin{equation}
  \label{eq:atm-model-1}
  K_{\atm}(\lambda, \hat{z}) =\sum_{j} X^{\rho_{j}}(\hat{z})
  \times k_{j}(\lambda).
\end{equation}
Here the different physical components $j$ are:
\begin{itemize}
  \item Rayleigh scattering, $k_{R}$,
  \item aerosol scattering, $k_{A}$,
  \item ozone absorption, $k_{\Oz}$,
  \item telluric absorption, $k_{\tell}$.
\end{itemize}
$X$ denotes airmass, and $\rho_{j}$ is an airmass correction exponent
(Beer-Lambert law in presence of saturation), with $\rho_{j} = 1$
\citep{Rufener86, Burke10} for all but the telluric component.

In the following subsections, we will present the different extinction
components as well as the time dependent part of the atmospheric
transmission, $\delta T_{i}$.

\subsection{Light scattering}
\label{sec:mie-scattering}

The treatment of light scattering depends on the ratio between the
scattering particle size and incident wavelength.  Light scattering by
particles of size comparable to the incident wavelength is a complicated
problem which has an exact solution only for a homogeneous sphere and a
given refractive index.  This solution was proposed by \cite{Mie08} to
study the properties of light scattering by aqueous suspensions of gold
beads.  A limiting case of this problem --- the Born approximation in
quantum mechanics --- is Rayleigh scattering, for which the size of the
particles is very small compared to the incident wavelength.  At the
other extreme, when the wavelength becomes much smaller than the size of
the scattering particles (like the water droplets or ice crystals in the
clouds), the scattering cross section becomes constant and of order of
the geometrical cross section. We cover these different cases in the
following subsections.

\subsubsection{Rayleigh scattering}
\label{sec:rayleigh-scattering}

As introduced above, Rayleigh scattering refers to scattering of light
by atoms or molecules (those of the atmosphere in our case) whose size
is much smaller than the incident wavelength.  For molecules in
hydrostatic equilibrium, extinction due to Rayleigh scattering can be
expressed as,
\begin{equation}
  \label{eq:rayleigh-1}
  k_{R}(\lambda, P, h) =
  \frac{2.5}{\ln(10)} \frac{\sigma(\lambda)\ P}{g(h)\ M},
\end{equation}
where $P$ is the atmospheric pressure at the site, $M$ is the molecular
mass of dry air, $g(h)$ is the equivalent acceleration of gravity at the
altitude $h$, and $\sigma(\lambda)$ represents the Rayleigh cross
section for dry air \citep{Bucholtz95, Breon98}.
Eq.~\eqref{eq:rayleigh-1} is a simplified equation of a more general
case including water vapor in the atmosphere, but it turns out that this
correction factor is negligible (of the order of $10^{-3}$) in this
analysis.  In this case, the variation in Rayleigh extinction at a given
observing site depends only on surface pressure, $P$.

\cite{Sissenwine62} and \cite{Bucholtz95} have tabulated values for
cross sections of the Rayleigh scattering for a wavelength range from
0.2 to 1~$\mu$m.  These data allowed \cite{Bucholtz95} to fit an
analytical model for the cross section:
\begin{equation}
  \label{eq:rayleigh-2}
  \sigma(\lambda) = A \lambda^{-\left(B + C \lambda + D/\lambda \right)},
\end{equation}
where $A$, $B$, $C$ and $D$ are numerical parameters.

\begin{figure}
  \centering
  \includegraphics[width=\columnwidth]{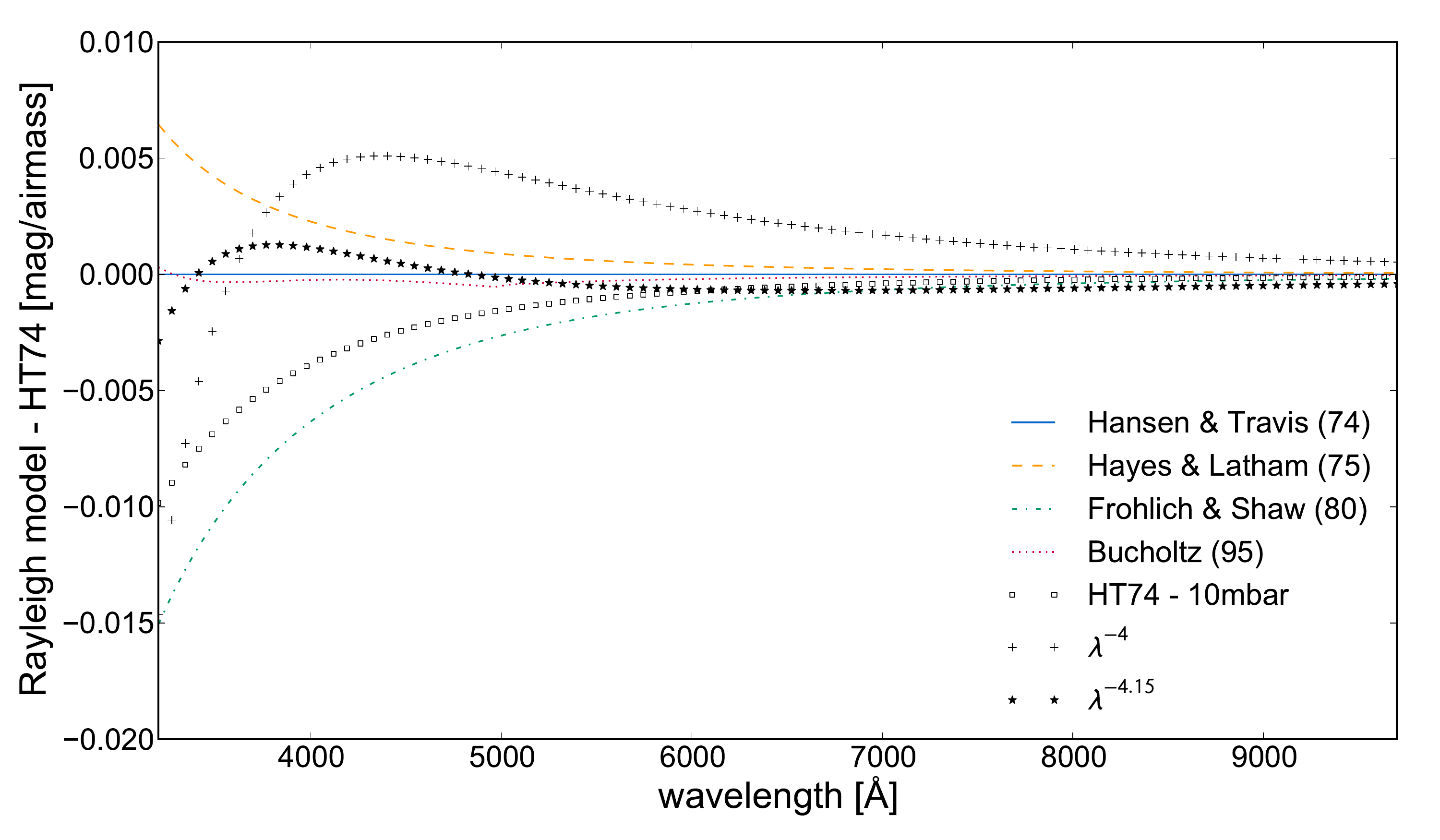}
  \caption{Comparison between different methods to model the Rayleigh
    scattering in the \snifs wavelength range from \cite{Hansen74}
    (blue), \cite{Hayes75} (yellow), \cite{Froehlich80} (green) and
    \cite{Bucholtz95} (red).  The comparison is made for the mean
    pressure at the Mauna Kea summit (616~mbar).  The discrepancies
    between the different models are less than 1.5\% over the wavelength
    range considered in this paper.}

  \label{fig:rayleigh-comparison}
\end{figure}

Fig.~\ref{fig:rayleigh-comparison} shows other numerical evaluations of
the Rayleigh scattering for a Mauna Kea mean pressure of 616~mbar,
including \cite{Hansen74}, \cite{Hayes75} and \cite{Froehlich80} whose
results are very similar to the model from \cite{Bucholtz95}.  Note that
all these methods are very close to the simplified formula
$\lambda^{-4+\epsilon}$, which is often found in the literature.  In our
analysis, the value of $\epsilon$ that best matches the Rayleigh
description at aforementioned 616~mbar is $-0.15$ (\cf
Fig.~\ref{fig:rayleigh-comparison}, solid black line).

Since we have a direct measurement of the surface pressure at the Mauna
Kea summit at the time of our observations, the Rayleigh scattering
component is not adjusted in our model. Rather, following common
practice in the atmospheric science community we directly use the
calculated Rayleigh extinction. For convenience we employ the
\cite{Hansen74} description, very close to the \cite{Bucholtz95}
expression.

\subsubsection{Aerosol scattering}
\label{sec:aerosols-scattering}

The monitoring of atmospheric aerosols is a fundamentally difficult
problem due to its varying composition and transport by winds over large
distances. For Mauna Kea we expect that aerosols of maritime origin,
essentially large sea-salt particles, will dominate
\citep{Dubovik02}. In that case, we can expect a low aerosol optical
depth given the elevation of Mauna Kea \citep{Smirnov01}.  Furthermore,
the strong temperature inversion layer between 2000 and 2500~m over the
island of Hawaii helps to keep a significant fraction of aerosols below
the summit\footnote{\url{http://www.esrl.noaa.gov/gmd/obop/mlo/}}. Major
volcanic eruptions can inject aerosols into the upper atmosphere,
affecting extinction \citep{Rufener86, Vernier11}.  Nearby Kilauea has
been active throughout the period of our observations, but its plume is
generally capped by the inversion layer and carried to the southwest,
keeping the plume well away from the Mauna Kea summit.

The particle sizes are of the order of the scattered wavelength for the
wavelength range (3000--10000~\AA) of our study.  According to
\cite{Angstrom29}, \cite{Angstrom64} and \cite{Young89}, and in
agreement with the Mie theory, the chromaticity of the aerosol
scattering is an inverse power law with wavelength:
\begin{equation}
  \label{eq:aerosol-1}
  k_{A}(\lambda) =  \tau \times (\lambda/1\;\mu\mathrm{m})^{-\aa}
\end{equation}
where $\tau$, the aerosol optical depth at 1~$\mu$m, and $\aa$, the \ang
exponent, are the two parameters to be adjusted.

According to \cite{Reimann92}, the value of the exponent $\aa$ varies
between $4$ and $-2$ for astronomical observations, depending on the
composition of the aerosol particles. We will see in
\S~\ref{sec:aerosol-mauna-loa} that the \ang exponent at the Mauna Kea
summit is confirmed to vary within these values, with a mean value close
to 1. While aerosol scattering may be spatially and temporally variable,
since it is anticipated to be weak at the altitude of Mauna Kea we begin
our study by assuming aerosol scattering is constant on the timescale of
a night. We defer the discussion of this hypothesis to
\S~\ref{sec:line-sight}.

\subsubsection{Water scattering in clouds and grey extinction}
\label{sec:water-scattering}

The water droplets and ice crystals in clouds also affect the
transmission of light through the atmosphere. To first approximation, it
can be assumed that the size of the constituents of the clouds are large
compared to the wavelength of the incident light (a cloud droplet
effective radius is of the order of at least 5~$\mu$m,
\citealt{Miles00}).  In this case, the dominant phenomenon is the
refraction inside water droplets. The extinction is then almost
independent of wavelength and can be considered achromatic.  We will
therefore refer to this extinction as ``grey extinction''.  In
\S~\ref{sec:short-variability} we will examine this approximation more
closely, and demonstrate is applicability for cloud conditions under
which useful astronomical observations are possible.

In our current framework, extinction by clouds is the only component
treated as variable on the time scale of a night (\cf
\S~\ref{sec:discussion}).  It is represented by the atmospheric
transmission variation parameter, $\delta T (\lambda, \hat{z}, t)$.
Being grey, there is no wavelength dependence, and we may write:
\begin{equation}
  \label{eq:water-1}
  \delta T(\lambda, \hat{z}, t) = \delta T (\hat{z}, t).
\end{equation}

\subsection{Molecular absorption}
\label{sec:molecular-absorption}

The molecular absorption bands and features in the atmosphere are
essentially due to water vapor, molecular oxygen, and ozone.  Nitrogen
dioxide exhibits broad absorption, but is too weak to affect
astronomical observations \citep{Orphal03}. The regions below 3200~\AA{}
and above 8700~\AA{} are especially afflicted, due to strong \Oz and
\HdO absorption, respectively.

\subsubsection{Ozone bands}
\label{sec:ozone-bands}

The ozone opacity due to the Hartley \& Huggins band \citep{Huggins90}
is responsible for the loss of atmospheric transmission below
3200~\AA{} and the Chappuis band \citep{Chappuis80} has a
non-negligible influence --- at the few percent level --- between
5000~\AA{} and 7000~\AA{}.

\begin{figure}
  \centering
  \includegraphics[width=\columnwidth]{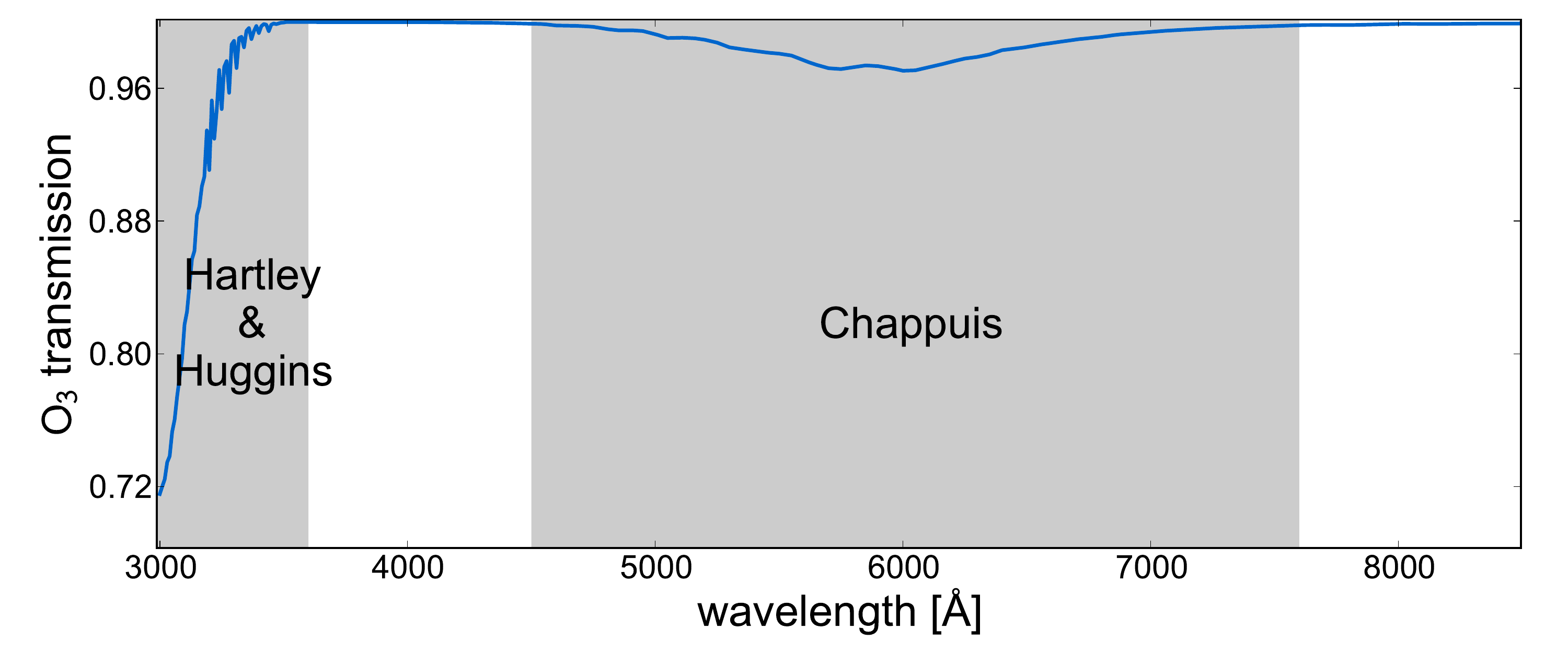}
  \caption{Ozone transmission template of the Hartley \& Huggins band in
    the UV and the Chappuis band at 6000~\AA.}
  \label{fig:ozone-template}
\end{figure}

In order to model the ozone absorption we are using a template (\cf
Fig.~\ref{fig:ozone-template}) for which we adjust the scale:
\begin{equation}
  \label{eq:ozone-1}
  k_{\Oz}(\lambda) = I_{\Oz} \times P_{\Oz}(\lambda)
\end{equation}
where $P_{\Oz}(\lambda)$ represents the ozone extinction template
computed using the MODTRAN library\footnote{\url{http://modtran.org/}}
and $I_{\Oz}$ is a scale factor, expressed in Dobson Units [DU].

\subsubsection{Telluric lines}
\label{sec:telluric-lines}

In contrast to the other extinction components, the telluric lines
affect only a few limited wavelength ranges.  The major features are
comprised of saturated narrow \Od lines, including the Fraunhofer ``A''
and ``B'' bands, deepest at 7594~\AA{} and 6867~\AA{} respectively, a
wide \HdO band beyond 9000~\AA{}, and \HdO absorption features in
several spectral regions between 6000~\AA{} and 9000~\AA. The weak O$_4$
features at 5322~\AA{} and 4773~\AA{} \citep{Newnham98} have been
neglected in our current treatment.  Table~\ref{tab:telluric-domains}
shows the wavelength ranges taken to be affected by telluric features
for purposes of \snf calibration.  These wavelength ranges were
determined from the high resolution telluric spectrum from Kitt Peak
National Observatory\footnote{\url{http://www.noao.edu/kpno/}}
\citep[KPNO,][]{Hinkle03}, matched to the \snifs resolution.

\begin{table}
  \caption{Wavelength ranges of telluric features, determined from the
    high resolution KPNO spectrum, matched to the resolution of \snifs
    (3.2~\AA{} for the red spectroscopic channel).}
  \label{tab:telluric-domains}
  \centering
  \begin{tabular}{lcc}
    \hline
    \hline
    Feature & Start & End \\
    & [\AA] & [\AA] \\
    \hline
    $\Od \gamma + \Of$ & 6270.2 & 6331.7 \\
    $\Od $ B           & 6862.1 & 6964.6 \\
    $\HdO$             & 7143.3 & 7398.2 \\
    $\Od $ A           & 7585.8 & 7703.0 \\
    $\HdO$             & 8083.9 & 8420.8 \\
    $\HdO$             & 8916.0 & 9929.8 \\
    \hline
  \end{tabular}
\end{table}

For the telluric lines, the airmass dependence, $\rho$, from
Eq.~\eqref{eq:atm-model-1} corresponds to a saturation parameter. Since
the telluric contributions are separated enough to be interpolated over,
it is possible to determine this saturation parameter, which according
to \citet{Wade88, Stubbs07} is approximately~0.5 for strongly saturated
lines and 1 for unsaturated lines.  In \S~\ref{sec:telluric-correction}
we will further discuss the value of $\rho$ for the telluric lines as
well as the method used to correct them.

\section{Nightly photometricity}
\label{sec:photometricity}

The photometricity of a night refers to its transmission stability in
time.  At the present time we are only able to separate nights with
clouds from those unlikely to have clouds. Besides clouds, the aerosol
and water absorption components of the extinction are the most
variable. However, because their variation is difficult to detect, we do
not currently include these components when assessing the photometricity
of a night.  Recall that our formalism and instrument capabilities allow
us to determine the extinction on both photometric and non-photometric
nights. Nights that are non-photometric simply use estimates of $\delta
T_{i}(\lambda, \hat{z}, t)$ obtained from the parallel imaging
channel. For this reason we can afford to be conservative in our
selection of photometric nights.

\begin{figure*}
  \centering
  \subfloat[Photometric night]{%
    \label{fig:skyprobe-phot}%
    \includegraphics[width=.85\textwidth]{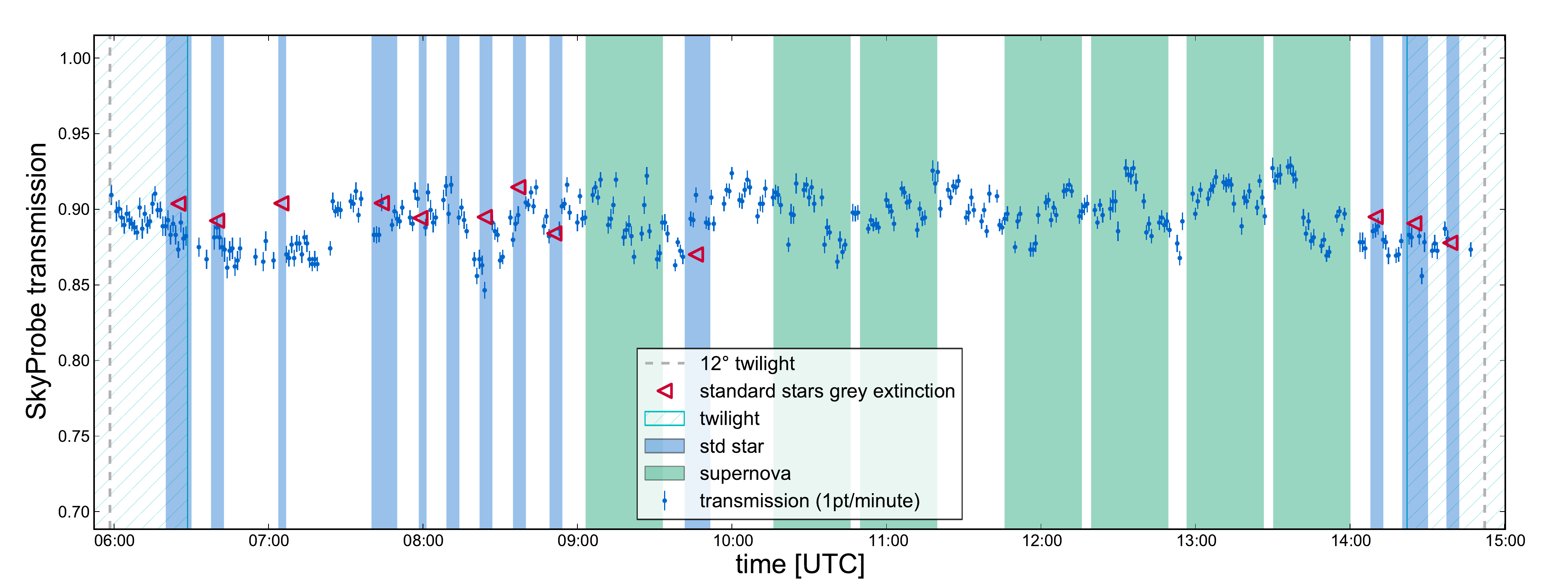}}\\
  \hspace{0mm}%
  \subfloat[Non-photometric night]{%
    \label{fig:skyprobe-nonphot}%
    \includegraphics[width=.85\textwidth]{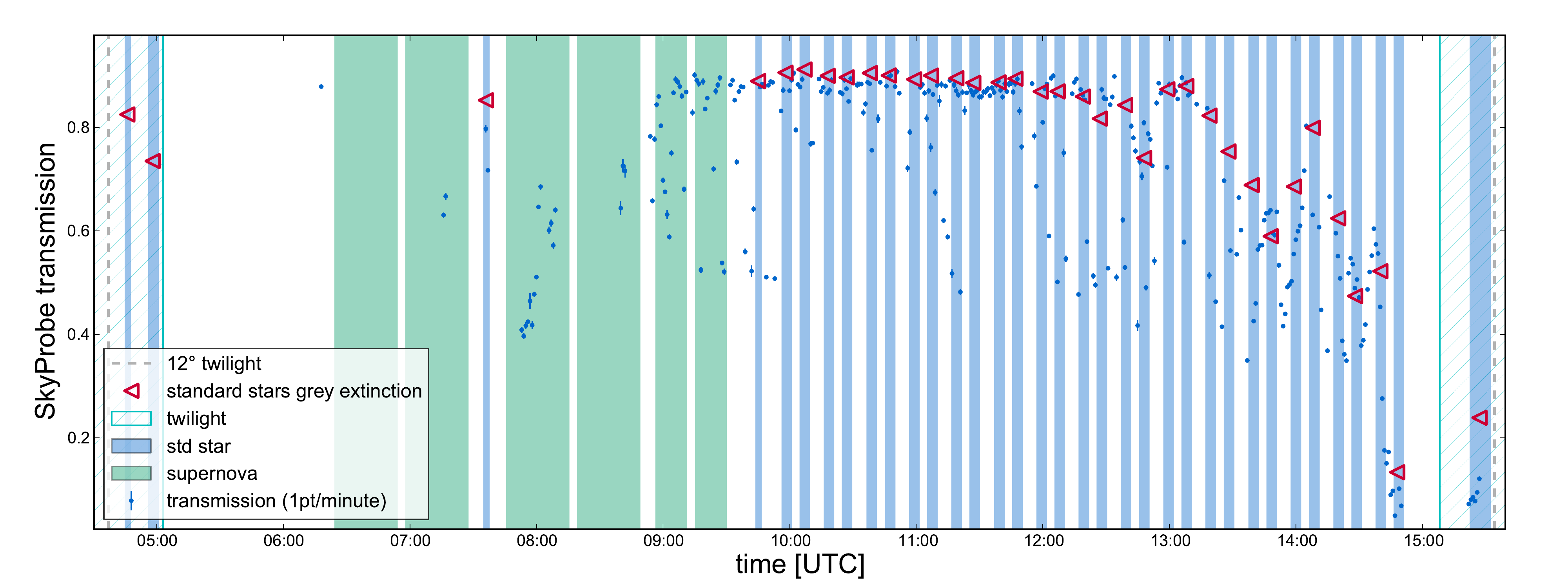}}%
  \caption{\protect\subref{fig:skyprobe-phot} \skp atmospheric
    transparency (blue points) and \snifs standard stars grey extinction
    term $\delta T$ (red triangles) for a photometric night and
    \protect\subref{fig:skyprobe-nonphot} a non-photometric night.}
  \label{fig:skyprobe}
\end{figure*}

\subsection{\skp}
\label{sec:skyprobe}

In order to obtain a reliable assessment of the sky transparency
stability, we use several available sources. These include \skp
\citep{Cuillandre02, Steinbring09}, photometry from the SNIFS parallel
imager, the brightness stability of the SNIFS guide stars, the scatter
of our standard stars about the best extinction solution, and knowledge
of technical issues such as dome slit misalignment.

Because of its high cadence and continuity across the sky, we begin with
measurements from
\skp\footnote{\url{http://www.cfht.hawaii.edu/Instruments/Elixir/skyprobe/home.html}}
\citep{Cuillandre02, Steinbring09}, a wide field camera mounted at the
CFHT and dedicated to real time atmospheric attenuation analysis.  Some
outlier cleaning of the \skp data is necessary since it includes
measurements taken when the telescope is slewing. There also is evidence
for occasional small but highly stable offsets between pointings,
suggestive of small systematics in the photometry reference catalog or
photometry technique employed. The robustness of such cleaning is
adversely affected on nights when CFHT slews frequently between fields.
We find that generally when the \skp data stream has an RMS greater than
3.5\% after cleaning the night is not likely to be photometric.  One
added limitation of our use of the \skp data is the possibility that
CFHT could miss the presence of clouds if only part of the sky is
affected throughout the night.

\subsection{Guide star}
\label{sec:guide-star}

Since the guiding video and resultant brightness measurements of SNIFS
guide stars are stored for all guided observations, the presence of
clouds in the SNIFS field can be ascertained directly. Some cleaning is
needed for these data as well, since cosmic ray hits or strong seeing
variations can produce measurable fluctuations in the guide star
photometry. The guide star video has a rate between 0.4 and 2~Hz, so the
data can be averaged over 30--60~sec intervals to achieve sensitivity at
the few percent level for most guide stars. As different guide stars
from the different targets observed over the course of a night have
different and unknown brightnesses, these data can only detect relative
instability over the interval of an observation but not between
observations and with poor sensitivity for short exposures.

\subsection{Tertiary reference stars}
\label{sec:photometric-std}

For fields that are visited numerous times, field stars in the \snifs
photometric channel can provide an estimate of the average attenuation
over the course of the parallel spectroscopic exposure. We refer to
these as a ``multi-filter ratio'' or MFR. The \snifs spectroscopic and
imaging channels cover adjacent regions of sky spanning just a few
arcminutes, and sit behind the same shutter. \cite{pereira08} found that
this relative photometry is accurate to $\sim 2$\%, except for the rare
cases where there are few suitable field stars. For long exposures of
supernovae there generally are enough field stars with high
signal-to-noise. Some standard star fields lack enough stars to ensure a
sufficient number of high signal-to-noise field reference stars for our
typical exposure times and under mildly cloudy conditions. Use of such
standards for our program has been phased out.

\subsection{Spectroscopic standard stars}
\label{sec:spectroscopic-std}

Finally, our formalism allows us to easily compute an initial instrument
calibration and extinction curve under the assumption that a night is
non-photometric.  The resulting values of $\delta T(\hat{z},t)$ can then
be used to detect the presence of clouds during the standard star
exposures themselves.

\subsection{Combined probes}
\label{sec:combined-probes}

Combining all this information, we estimate the photometricity of the
night for the targets observed by SNIFS.  It is important to consider
the noise floor for each source to avoid rejecting too many photometric
nights. By examining the distribution of each photometricity source we
are able to define two thresholds, ``intermediate'' and
``non-photometric''. For our purposes in this paper, non-photometric
nights are defined as having at least one source above a non-photometric
threshold, or all sources above the intermediate thresholds.  The
threshold values are listed in Table~\ref{tab:photo-cuts}.

\begin{table}
  \caption{Root Mean Square thresholds (in \%) for the different
    photometricity source distributions. For values intermediate between the
    photometric and non-photometric thresholds, a combination of indicators
    is used to ascertain the temporal stability of the atmospheric transmission.}
  \label{tab:photo-cuts}
  \centering
  \begin{tabular}{lcc}
    \hline
    \hline
    Sources & Photometric & Non-photometric \\
    \hline
    \skp & $<2.5$ & $>3.5$ \\
    \snifs guide star & $<2.5$ & $>5$ \\
    \snifs photometry & $<2.5$ & $>4$ \\
    \snifs standard stars & $<2.5$ & $>4$ \\
    \hline
  \end{tabular}
\end{table}

\begin{figure}
  \centering
  \includegraphics[width=\columnwidth]{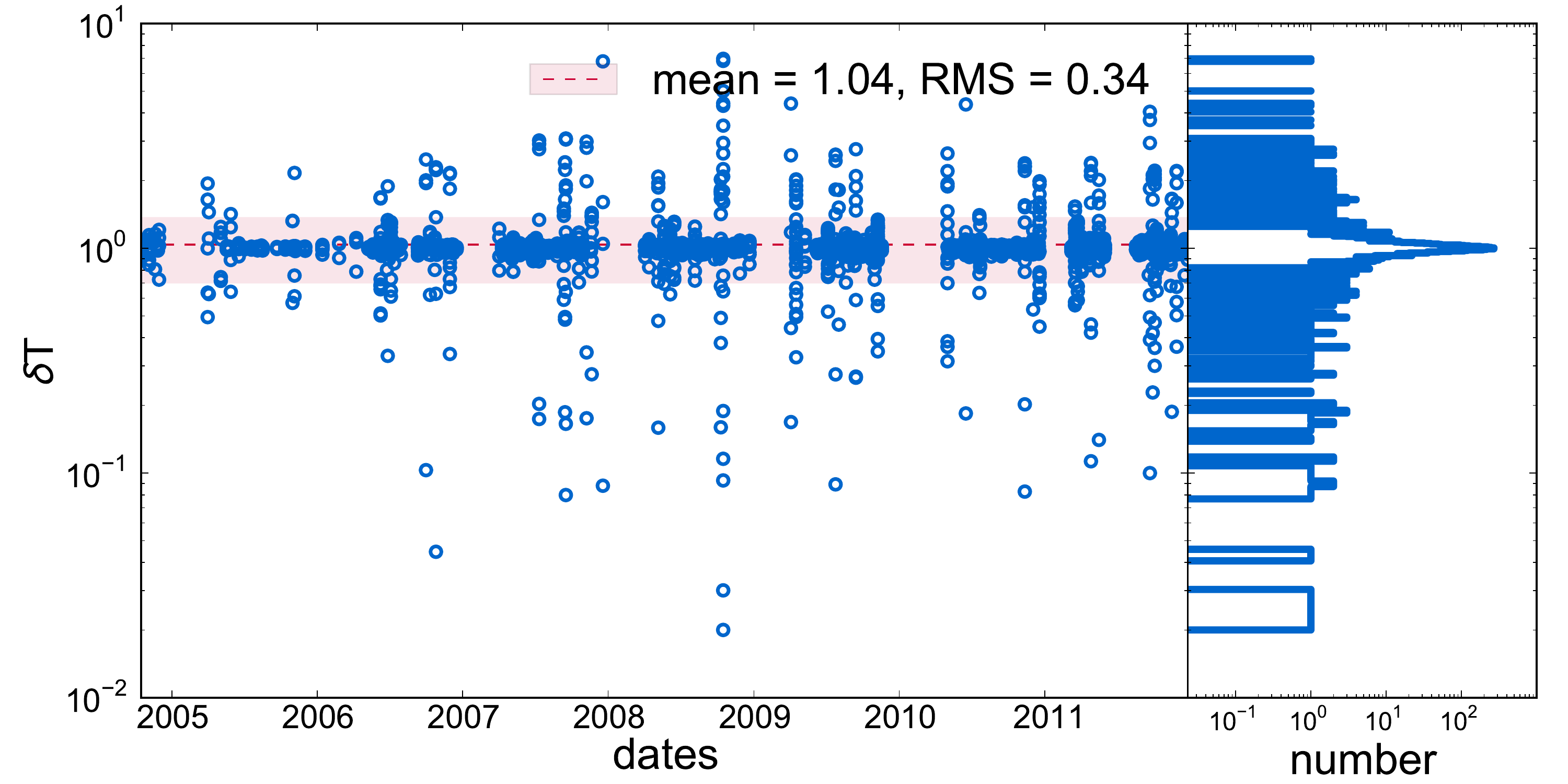}
  \caption{Distribution and histogram of the grey transmission
    parameter $\delta T$ for each observed standard star during
    non-photometric conditions. Note that many non-photometric nights have
    only thin clouds, and therefore $\delta T \sim 1$.}
  \label{fig:grey-histo}
\end{figure}

Examples of the \skp transmission are shown in
Fig.~\ref{fig:skyprobe-phot} for a photometric night and
Fig.~\ref{fig:skyprobe-nonphot} for a non-photometric night.  The red
triangles represent the grey extinction term, $\delta T_{i}(\hat{z},t)$,
for each standard star $i$ observed during the night. The pattern of the
grey extinction seen in Fig.~\ref{fig:skyprobe-nonphot} follows that of
the independent \skp data (blue points). This confirms that the
parameter $\delta T(\hat{z},t)$ is able to track atmospheric attenuation
by clouds (\cf Fig.~\ref{fig:grey-histo} to see the distribution of
$\delta T(\hat{z},t)$ in non-photometric conditions). We estimate that
$\sim 35\%$ of the \snf nights were photometric according to these
transmission stability cuts. This value is lower than values of 45--76\%
that have been reported elsewhere (\citealt{Schock09} \& ESO Search for
Potential Astronomical Sites,
ESPAS\footnote{\url{http://www.eso.org/gen-fac/pubs/astclim/espas/espas_reports/ESPAS-MaunaKea.pdf}}),
however as noted earlier, for our purposes in this paper we wish to set
conservative photometricity criteria.

\section{Extinction from standard star observations}
\label{sec:results}

In the case of standard star observations, $S^{\star}(\lambda, t)$
is known \emph{a priori} as a tabulated reference, $\overline{S}(\lambda)$, and
thus the term on the left side of Eq~\eqref{eq:formalism-7} becomes a
known quantity.  To solve for the extinction and instrument
calibration we begin by constructing a conventional~$\chi^{2}$:
\begin{equation}
  \label{eq:application-1}
  \chi^{2} = \sum_{i}
  \mathcal{R}_{i} \cdot V_{i}^{-1} \cdot \mathcal{R}_{i}^{\text{T}},
\end{equation}
where the index $i$ stands for each individual standard star, $V_{i}$
is the covariance matrix (described below) and $\mathcal{R}_{i}$
represents the residuals, given by,
\begin{equation}
  \label{eq:application-1bis}
  \mathcal{R}_{i} = \log C(\lambda) -
  0.4 \times K_{\atm}(\lambda, \hat{z}) +
  \log \delta T_{i}(\hat{z}, t) -
  \log\frac{S_{i}(\lambda, \hat{z}, t)}{\overline{S}_{i}(\lambda)}.
\end{equation}
$K_{\atm}(\lambda, \hat{z})$ represents the parametrization detailed in
\S~\ref{sec:atmospheric-extinction} for the Rayleigh, aerosols, ozone
and telluric extinction components. The only adjustable parameters
are:
\begin{itemize}
\item $C(\lambda)$, the instrument calibration (\cf Eq.~\ref{eq:formalism-2}),
\item $\aa$ and $\tau$, the aerosol \ang exponent and optical
  depth,
\item $I_{\Oz}$, the ozone template scale factor,
\item $\delta T_{i}(\hat{z}, t)$, the transmission variation for each
  standard star $i$.
\end{itemize}
$k_{R}(\lambda)$ is not adjusted since it depends only on the surface
pressure $P$, which is known \emph{a priori}.  The instrument
calibration, $C(\lambda)$, and the atmospheric extinction,
$K_{\atm}(\lambda, \hat{z})$, are spectrally smooth, so we do not
constrain the model at full resolution. Rather, we employ a coarser
spectral grid, consisting of ``meta-slices''; we construct meta-slices
for each of the two \snifs channels, with meta-slice widths between 100
and 150~\AA{} depending on the channel. The telluric line scale factors
must also be determined, but we will see in the following section that
this can be accomplished in a separate step.

$V_{i}$ in Eq.~\ref{eq:application-1} is the covariance matrix between
all meta-slice wavelengths of standard star $i$ for a given night; this
assumes the covariance between standards is zero. To build the
covariance matrix of each standard star, we first include the
statistical covariance issued from the point-source PSF-extraction
procedure. A constant is then added to the whole matrix, representing a
3\% correlated error between all meta-slices for a given standard star
observation. This is added as a way to approximate our empirically
determined per-object extraction error.

As described in \S~\ref{sec:priors}, when solving for the instrument
calibration and extinction we will modify Eq.~\ref{eq:application-1}
to include Bayesian priors on some atmospheric parameters.

\subsection{Telluric correction}
\label{sec:telluric-correction}

The telluric lines affect only limited wavelength regions, whereas the
other atmospheric extinction components are continuous (aerosol and
Rayleigh scattering) or very broad (ozone). Although it is possible to
fit the telluric lines at the same time as the extinction continuum, we
chose to do it in a separate step both for historical implementation
reasons and to allow a telluric correction when the flux calibration is
not needed. As a consequence, with our approach the telluric wavelength
regions can be either avoided for the study of the atmospheric
extinction continuum or corrected separately using standard stars to
determine an average correction spectrum per night. In the latter case we
separate $K_{\atm}(\lambda, \hat{z})$ as follows:
\begin{equation}
  \label{eq:telluric-1}
  K_{\atm}(\lambda, \hat{z}) = \underbrace{%
    X(\hat{z}) \times
    \left(
      k_{R}(\lambda) + k_{A}(\lambda) + k_{\Oz}(\lambda)
    \right)
  }_{K(\lambda, \hat{z})} +  X^{\rho}(\hat{z}) \times k_{\tell}(\lambda).
\end{equation}
For a standard star $i$ on a given night,
\begin{equation}
  \label{eq:telluric-2}
  \mathscr{C}^{\star}_{i}(\lambda, \hat{z}, t) =
  \frac{S_{i}(\lambda, \hat{z}, t)}{\overline{S}_{i}(\lambda)} =
  \delta T_{i}(\hat{z}, t) \times C(\lambda) \times
  10^{-0.4 \times K(\lambda, \hat{z})},
\end{equation}
behaves as a smooth function of wavelength, which can be reasonably well
modeled with a spline $\mathscr{C}(\lambda, \hat{z}, t)$, outside the
telluric lines regions. Inserting this into Eq.~\eqref{eq:formalism-7}
for standard stars gives:
\begin{equation}
  \label{eq:telluric-3}
  \log  \frac{S_{i} (\lambda, \hat{z}, t)}{\overline{S}_{i}(\lambda)}  =
  \log \mathscr{C}_{i}(\lambda, \hat{z}, t) -
  0.4 \times X_{i}^{\rho}(\hat{z}) \times k_{\oplus}(\lambda).
\end{equation}

As shown in \S~\ref{sec:atmospheric-extinction}, the amplitude of the
telluric lines is proportional to the factor $X^{\rho}(\hat{z})$.
Taking the logarithm of Eq.~\eqref{eq:telluric-3}, we obtain a linear
expression with respect to the logarithm of the airmass, $\log
X(\hat{z})$, allowing us to fit for the saturation factor $\rho$ and the
telluric extinction $k_{\oplus}(\lambda)$:
\begin{equation}
  \label{eq:telluric-4}
  \log \left( -2.5 \times
    \log \frac{S_{i}(\lambda, \hat{z}, t)/\overline{S}_{i}(\lambda)}{%
      \mathscr{C}_{i}(\lambda, \hat{z}, t)}
  \right) = \log k_{\oplus}(\lambda) + \rho \times \log X_{i}(\hat{z}).
\end{equation}

\begin{figure}
  \centering
  \includegraphics[width=\columnwidth]{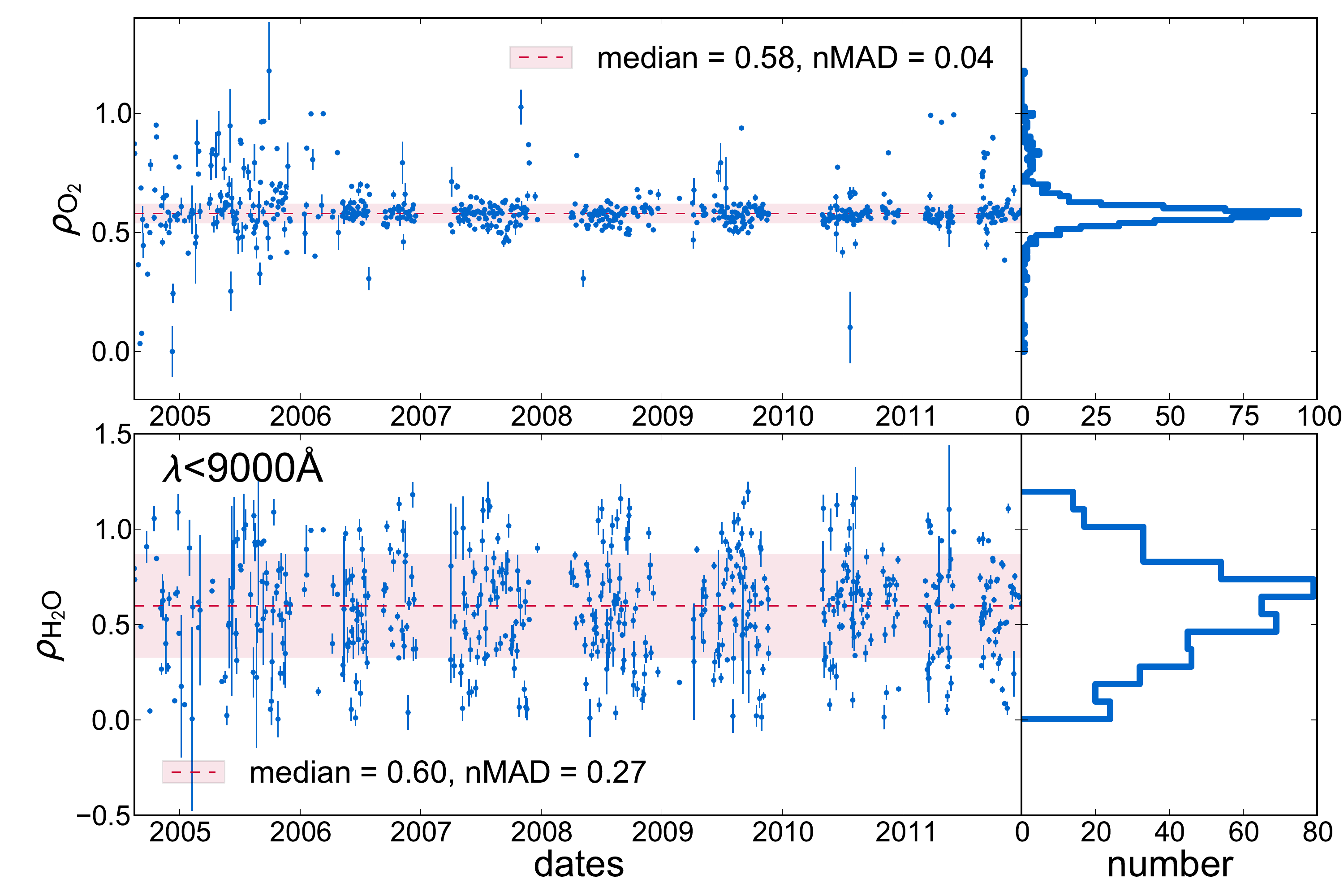}
  \caption{Distributions of the saturation parameters, $\rho_{\Od}$ and
    $\rho_{\HdO}$, for each night of the \snf data set. The water lines
    above 9000~\AA{} are treated separately due to the difficulties
    encountered in the correction of the ``strong'' \HdO telluric
    band. One reason we chose to fix the water saturation parameter is
    that some fits gave unphysical values (\eg $\rho_{\HdO} > 1$).}
  \label{fig:telluric-saturation}
\end{figure}

Optically thin absorption produces attenuation proportional to the
airmass ($\rho=1$) whereas highly saturated lines are expected to have
equivalent widths growing as the square root of the airmass
($\rho=0.5$). According to the observations performed by \cite{Wade88}
for airmasses from 1 to 2, the saturated Fraunhofer ``A'' and ``B''
lines and the water lines have an airmass dependence $\rho \simeq 0.6$.
We repeated their approach and fit for $\rho_{\Od}$ and $\rho_{\HdO}$
for each observation night. The saturation distributions are shown in
Fig.~\ref{fig:telluric-saturation}. We find a median value of 0.58 for
$\rho_{\Od}$, and find a normalized Median Absolute Deviation (nMAD) of
0.03. As for the \HdO lines, since the water content in the atmosphere
is variable, so is the saturation parameter $\rho_{\HdO}$. We found
median values of $0.60\pm0.27$ and $0.35\pm0.37$ for water regions below
and above $\approx 9000$~\AA{}, respectively.

\begin{figure*}
  \centering
  \includegraphics[width=\textwidth]{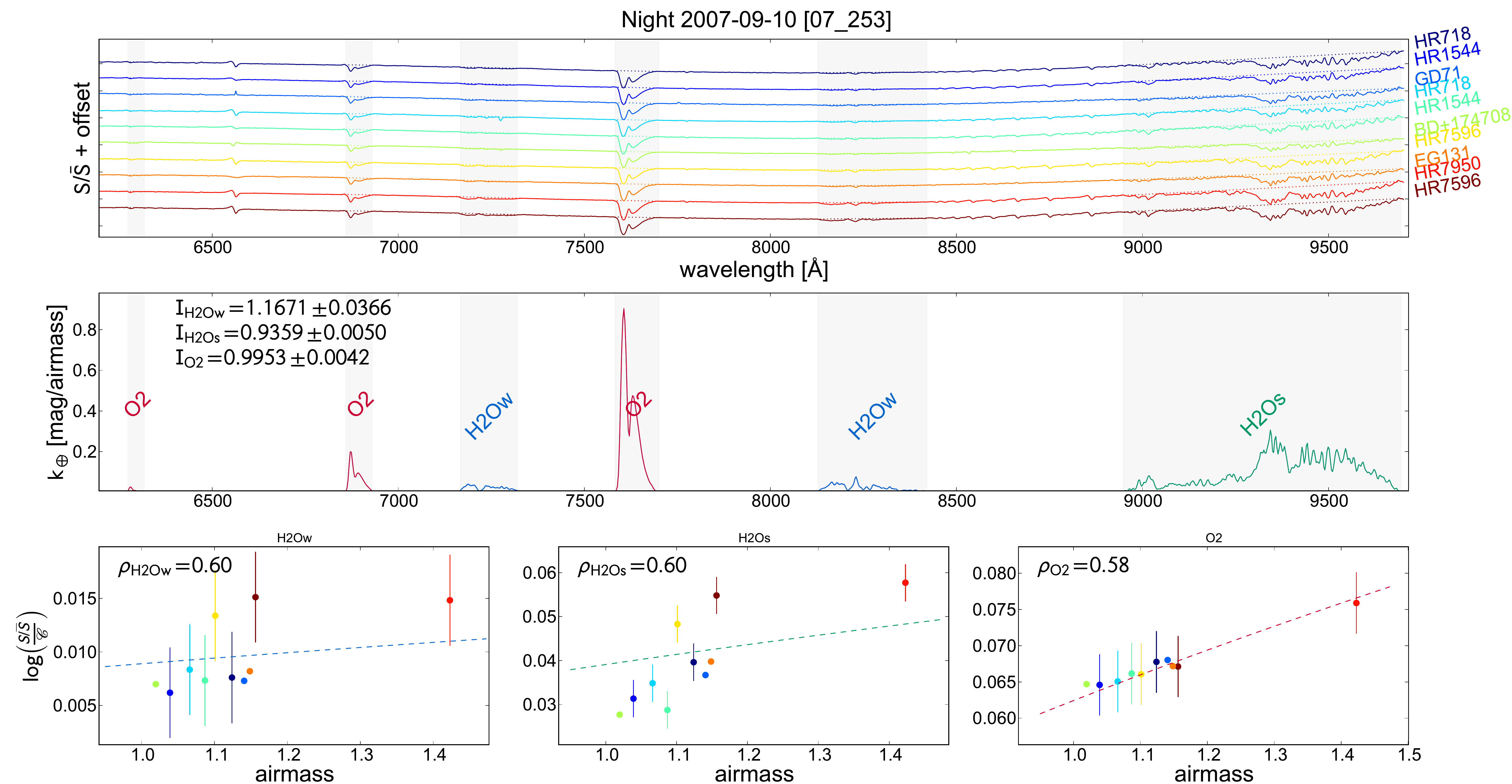}
  \caption{Example of a telluric correction spectrum computation for a
    typical night.  Top panel: All the reference normalized standard
    star spectra $S/\overline{S}$ of the night and their continua
    $\mathscr{C}$ (dashed lines).  Middle panel: Adjusted telluric
    template intensities as function of their
    saturation. Bottom panels: Dependence with respect to
      airmass of $\log S/\overline{S} - \log \mathscr{C}$ integrated
      over each group of lines (shaded areas in middle panel). The
      dashed lines represent the airmass dependence with a
      \emph{fixed} saturation exponent.}
  \label{fig:telluric-computation}
\end{figure*}

The latter result for the strong \HdO absorption region (8916--9929~\AA)
is treated independently from the region below $\approx 9000$~\AA{}
since it cannot be measured very well. This is because the wavelength
range of this particular band extends beyond the \snifs wavelength
range, which makes the estimation of the spline fit for $\mathscr{C}$
unreliable. As we can see in Fig.~\ref{fig:telluric-computation}, the
``strong'' water telluric band is difficult to correct in these
conditions. Nevertheless, by fixing the saturation parameter, a
correction accurate to 5--10\% can still be obtained.

Subsequently, and in order to improve the fit, we choose to neglect the
variations of the saturation and we fix $\rho_{\Od} = 0.58$ for the
molecular oxygen absorption and to $\rho_{\HdO} = 0.6$ for the water
vapor absorption \citep[as did][]{Wade88}. Neglecting the saturation
variations for the water bands has no perceptible negative impact on the
quality of the telluric correction since there is still a degree of
freedom from the telluric scale factor.

After removing the pseudo-continua, $\mathscr{C}(\lambda, \hat{z}, t)$,
from the spectra, linear fits of the intensity are performed for each
group of lines of the \snf derived telluric template (\ie \Od and \HdO)
with their respective saturation parameter fixed.

The results of the linear fit for $k_{\oplus}(\lambda)$ from a given
night are shown on Fig.~\ref{fig:telluric-computation}. The first panel
shows each observed spectrum divided by its corresponding intrinsic
spectrum (solid lines), together with the pseudo-continuum
$\mathscr{C}(\lambda, \hat{z}, t)$ (dashed lines) which is a global fit
for $S/\overline{S}$ in which the telluric regions are
excluded. The resulting telluric correction template is presented in the
second panel.  The linear fits for each group of lines are shown in the
three bottom panels.  We use $\mathrm{H_{2}Os}$ to denote the strong
\HdO telluric lines redward of 9000~\AA{}, and $\mathrm{H_{2}Ow}$ to
denote the weaker \HdO telluric lines blueward of this.  In the middle
bottom panel we see that the $\mathrm{H_{2}Os}$ correction is poorly
constrained due to the large scatter, as expected.  (We eventually
expect to improve this situation by extending the spectral extraction
beyond 1~$\mu$m.)

The current accuracy of the telluric correction is generally at the same
level as the noise of the spectra, and thus sufficient for rigorous
spectral analysis, including studies of spectral indicators
\citep{Bailey09, Chotard11}. For the oxygen lines, which are very
narrow, some wiggles ($\sim 2$\% peak to peak fluctuations) can remain
after the correction in a few cases due to small random wavelength
calibration offsets (of the order of $\sim 0.1$~\AA)
between the spectra and the template.

\begin{figure*}
  \centering
  \includegraphics[width=1.01\textwidth]{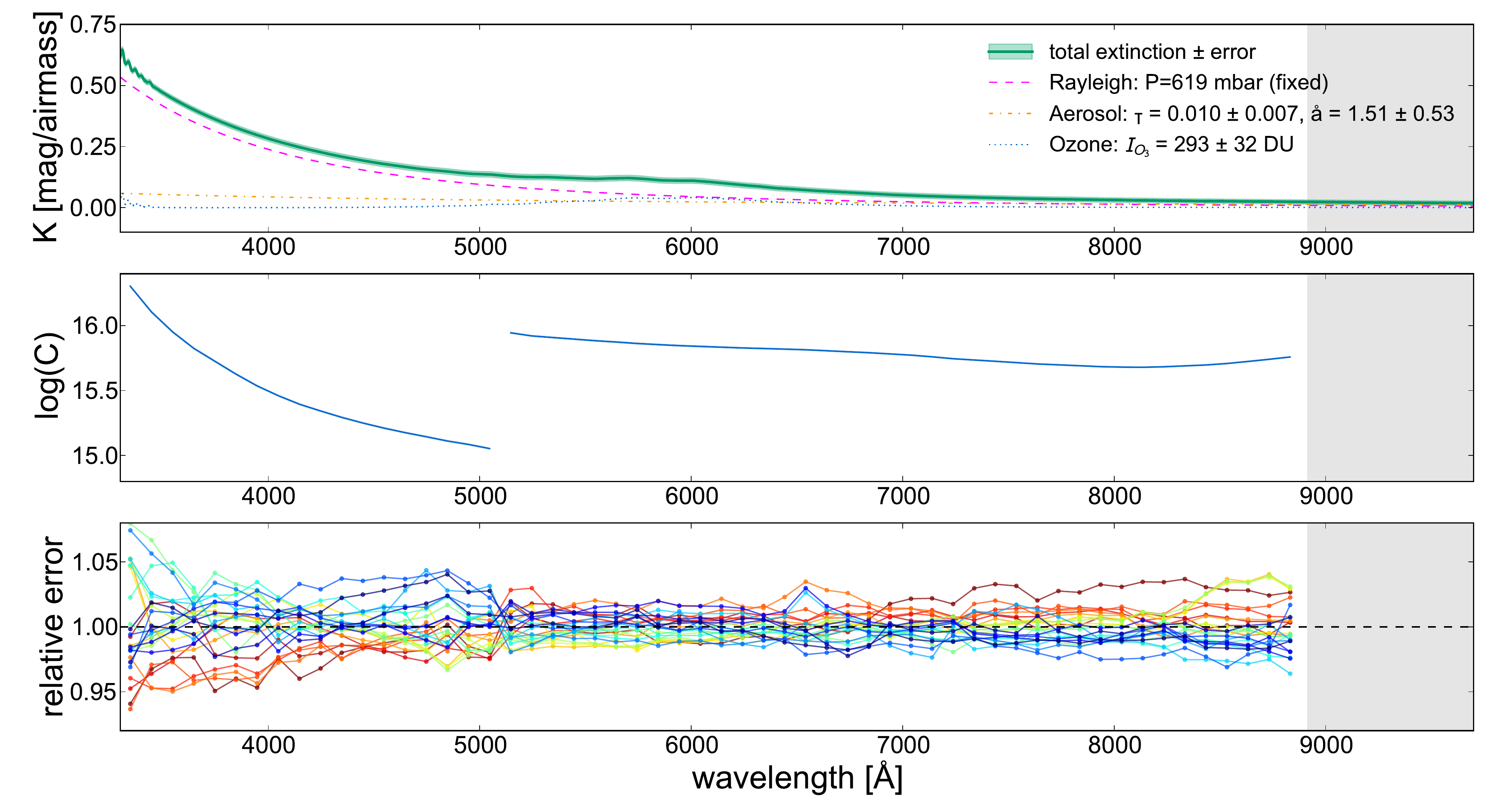}
  \caption{Upper panel: Atmospheric extinction, $K(\lambda)$ (solid
    line), for a given non-photometric night (2010-06-28); \ie the sum
    of its physical components, Rayleigh scattering $k_{R}(\lambda)$
    (dashed line), aerosol scattering $k_{A}(\lambda)$ (dash-dot line)
    and ozone absorption $k_{\Oz}(\lambda)$ (dotted line).  Middle
    panel: instrument calibration $C(\lambda)$; this includes not only
    the overall throughput, but also scaling by the spectral flat field.
    Bottom panel: Relative error of the standard star fluxes after the
    fit (\ie the difference between the individual standard star flux
    solutions and $\log C(\lambda)$). The grey bands in all three panels
    indicate meta-slices that are affected by the ``strong'' \HdO
    telluric features. This wavelength region is not used when fitting
    for the other physical components.}
  \label{fig:comp-extinction}
\end{figure*}

\begin{figure*}
  \centering
  \includegraphics[width=1.01\textwidth]{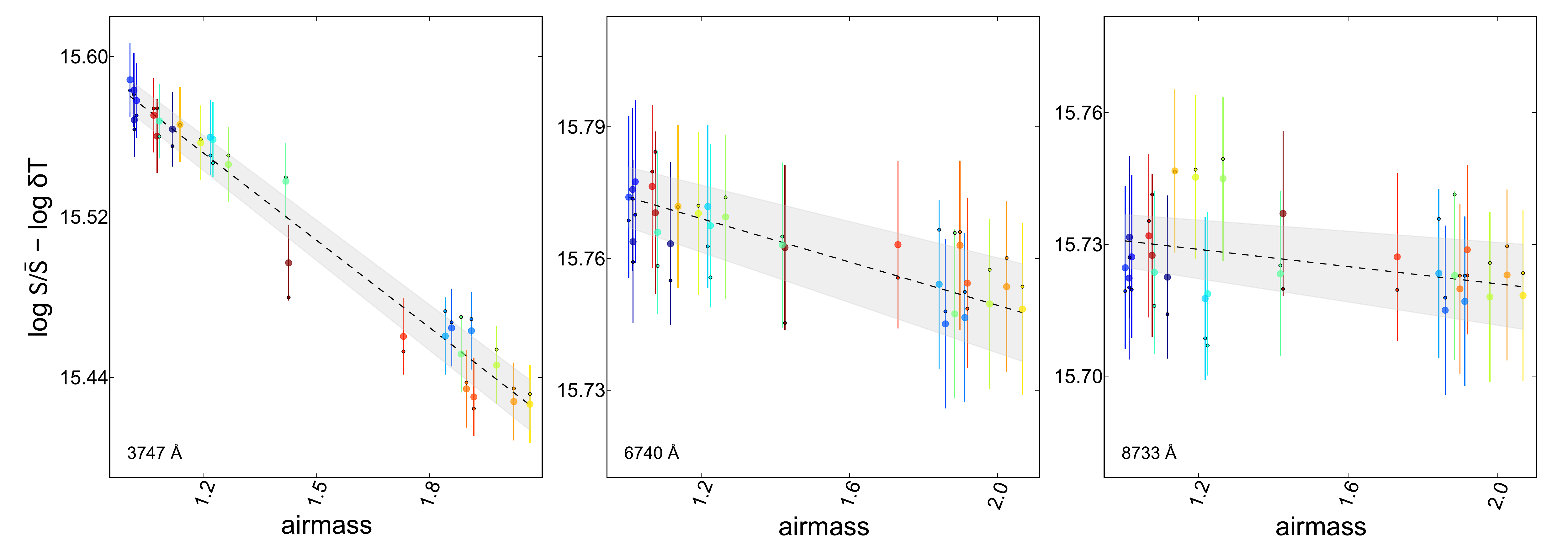}
  \caption{Linear fit of $\log S/\overline{S} - \log\delta T$ with
    respect to the airmass for three different wavelengths (indicated by
    dashed lines in the top panel), where each point represents a star
    with (large points with error bars) or without (small points) grey
    extinction correction.  The dashed lines represent the linear fits
    and the grey bands their corresponding error.}
  \label{fig:extinction-airmass}
\end{figure*}

\subsection{Nightly atmospheric extinction solutions}
\label{sec:multi-stand-approach}

\subsubsection{Introduction of Bayesian priors}
\label{sec:priors}

Outside  of the  telluric lines,  or  after correction  of the  telluric
features, Eq.~\eqref{eq:formalism-7} for standard stars becomes:
\begin{equation}
  \label{eq:multi-standard-1}
  \log \frac{S_{i} (\lambda, \hat{z}, t)}{\overline{S}_{i} (\lambda)}
  = \log C(\lambda) - 0.4 \times X_{i}(\hat{z}) \times K(\lambda) +
  \log \delta T_{i}(\hat{z}, t).
\end{equation}

Under these conditions, a degeneracy appears in
Eq.~\eqref{eq:multi-standard-1} between the instrument calibration $\log
C(\lambda)$ and the average of the grey extinction parameters $\langle
\log \delta T_{i} \rangle$. This means that the value of $C(\lambda)$
and the geometric average of the $\delta T_{i}$ values will only be
relative measurements during non-photometric nights.  The absolute scale
of the $\delta T_{i}$ can be determined independently though using MFRs
obtained by secondary stars from the \snifs photometric channel.  Such
MFRs are not needed to compute the atmospheric extinction, but we refer
the interested reader to \cite{pereira08} and \cite{Scalzo10} for
further description of this particular step of the \snifs flux
calibration process.

This degeneracy can be lifted during the fit by setting the mean value
of the grey extinction parameter to an arbitrary value.  For a
photometric night there should be no need for a grey extinction term
($\delta T_{i} \equiv 1$), thus $\langle \log \delta T_{i} \rangle$
would be zero if such a term were included.  To ease comparisons between
photometric and non-photometric nights, or the effects of changing the
photometric status of a night, we therefore set the mean value of the
cloud transmission to 1 on non-photometric nights. Since often such
nights have only thin cirrus or clear periods, this can provide
meaningful information on cloud transmission for other types of
studies. Note that mathematically the 3\% correlated error put into the
covariance matrix, $V$, can be traded off against the grey extinction
term on non-photometric nights. However, this has no effect on the
photometric solution.

In addition, in order to ensure that all components behave physically,
we chose to also apply priors to the aerosol and ozone parameters. This
choice is motivated by the fact that degeneracies can appear between the
physical components in some wavelength ranges resulting in negative
extinction or poor numerical convergence (For example, on
non-photometric nights there is a degeneracy between $\delta T$ and
$\tau \sim \lambda^{0}$).  All priors are Gaussian, but we chose to set
a prior on the logarithm of the aerosol optical depth since this
provides a better match to the aerosol optical depth distribution (\cf
Fig.~\ref{fig:optical-depth}) at the atmospheric observatory on nearby
Mauna Loa. All priors are summarized in Table~\ref{tab:priors}, and they
are implemented by adding a penalty function to the $\chi^{2}$ of
Eq.~\eqref{eq:application-1}:
\begin{equation}
  \label{eq:multi-standard-2}
  \chi^{2}_{\mathrm{total}} = \chi^{2} + \Psi^{2}
\end{equation}
where the priors on the atmospheric extinction shape are
  encapsulated in the penalty function,
\begin{equation}
  \label{eq:multi-standard-4}
  \Psi^{2} = \left( \frac{I_{\Oz} - {I^{\star}_{\Oz}}}{\sigma^{\star}_{I_{\Oz}}} \right)^{2} +
  \left( \frac{\aa - \aa^{\star}}{\sigma^{\star}_{\aa}} \right)^{2} +
  \left( \frac{\ln ( \tau / \tau^{\star} ) }{\varsigma^{\star}_{\tau}} \right)^{2}.
\end{equation}
The extinction for a given night is obtained by minimizing
Eq.~\eqref{eq:multi-standard-2} for all std stars $i$, and all
(meta-slices) wavelengths $\lambda$.

\begin{table}
  \caption{List of priors used in the fit for the
    atmospheric extinction and their corresponding errors.
    Their goal is to ensure physical behavior of the
    extinction components.}
  \label{tab:priors}
  \centering
  \begin{tabular}{cr@{ $\pm$ }lc}
    \hline
    \hline
    Prior & \multicolumn{2}{c}{Value} & Scaling \\
    \hline
    $I^{\star}_{\Oz}$ & 260 & 50~DU& linear \\
    $\tau^{\star}$ & 0.007 & 80\% & logarithmic \\
    $\aa^{\star}$ & 1 & 3 & linear \\
    \hline
  \end{tabular}
\end{table}

\subsubsection{Maximum-likelihood fitting}
\label{sec:bayesian-priors-and-fit}

The total $\chi^{2}$ (Eq.~\eqref{eq:multi-standard-2}), that includes
the Bayesian penalty function $\Psi^{2}$ is minimized. Consequently, the
resulting error on the parameters is computed by inverting the
covariance matrix of the full function.

\begin{figure}
  \centering
  \includegraphics[width=\columnwidth]{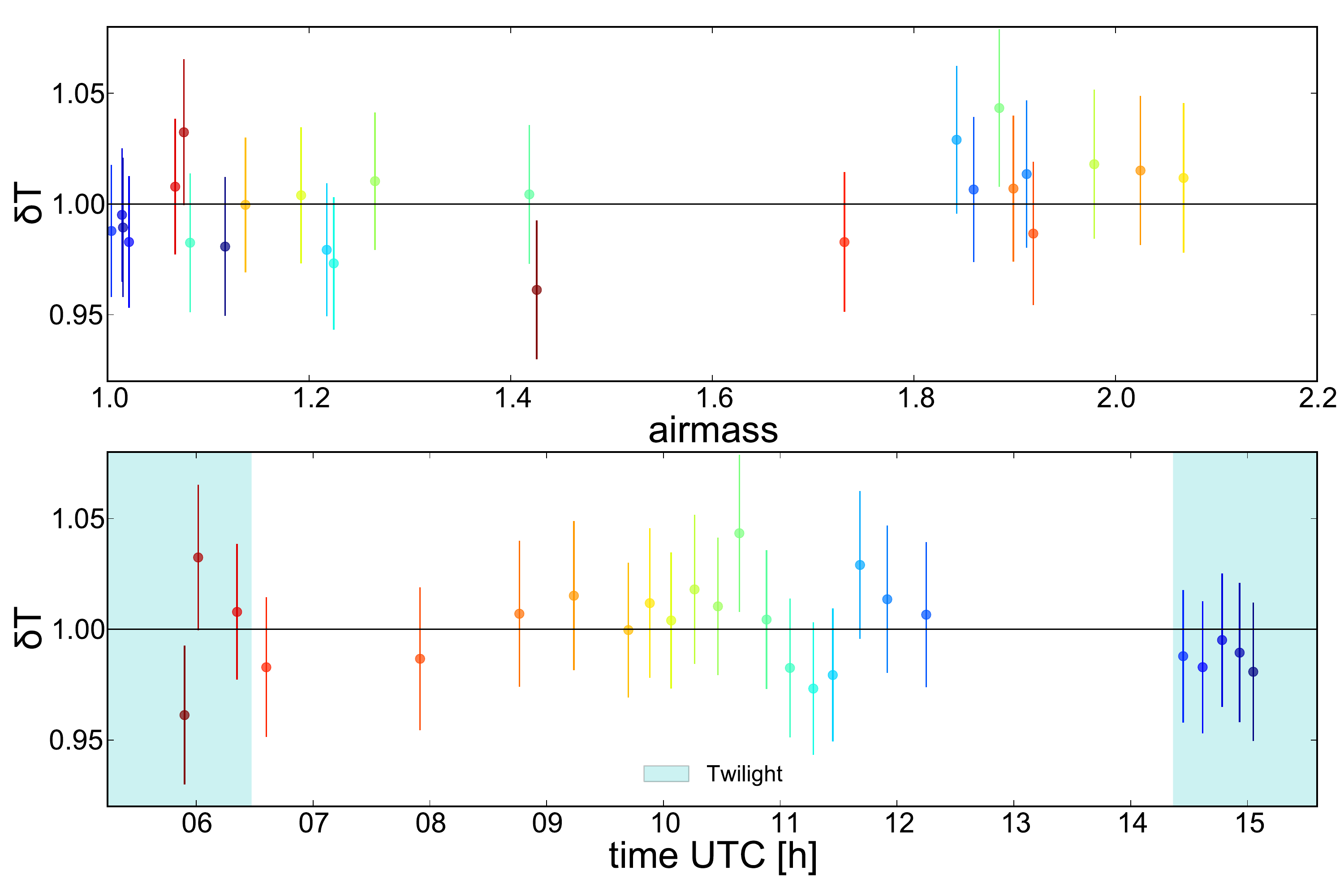}
  \caption{The grey extinction distribution, $\delta T(\hat{z}, t)$, for
    a slightly non-photometric night (2010-06-28). The top panel shows
    $\delta T(\hat{z}, t)$ as a function of airmass while the bottom
    panel shows $\delta T(\hat{z}, t)$ as a function of time. Since the
    night is non-photometric most of the scatter is due to clouds, but
    there is some scatter due to achromatic extraction errors. Recall
    that we normalize by the mean of $\delta T(\hat{z}, t)$ rather than
    the largest (fewest clouds) value.}
  \label{fig:extinction-alphas}
\end{figure}

\begin{figure*}
  \centering
  \includegraphics[width=.65\textwidth]{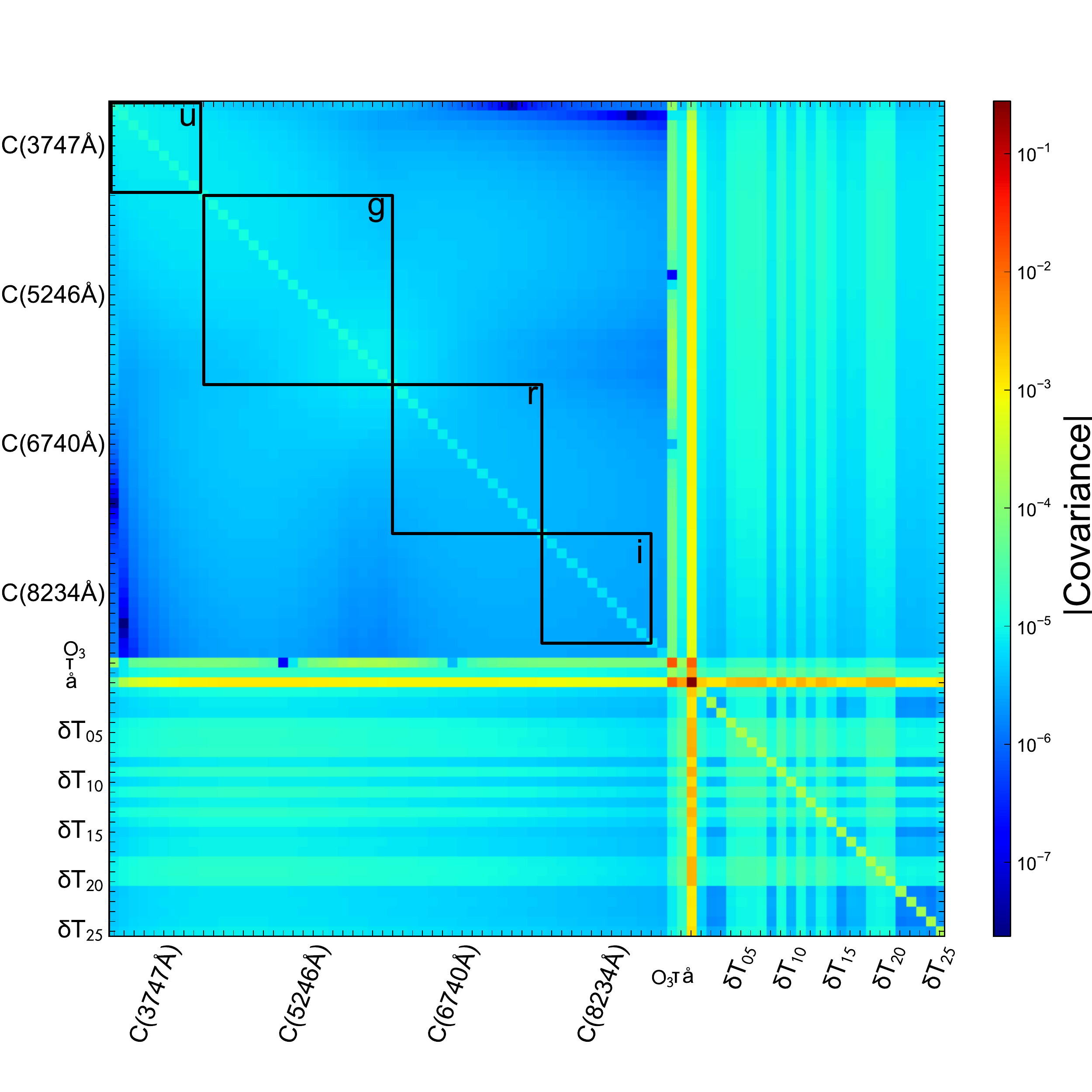}
  \caption{Resulting covariance matrix of the parameters for a given
    non-photometric night (2010-06-28). The represented parameters are
    the flux solution $C(\lambda)$ at every meta-slice wavelength, the
    atmospheric \textit{normalized model parameters} $I_{\Oz}$, $\tau$
    and $\aa$ as well as the grey extinction $\delta T$ per star. SDSS
    filters u, g, r and i are also represented (black squares) over the
    flux solution wavelengths.}
  \label{fig:cov-matrix}
\end{figure*}

The different contributions to the resulting extinction fit for a
specific night are illustrated in Fig.~\ref{fig:comp-extinction}, along
with the instrument calibration. The noticeable larger scatter in the
blue arm can be explained by the PSF model used for point-source
extraction which is presumably less accurate in the blue than in the
red, because of a stronger chromaticity; this is even more evident for
the short exposure PSF model used for bright standard stars, due to the
complex PSF in short exposures. Overall, increased errors in
point-source extracted spectra in the blue will contribute to a larger
scatter in the flux calibration residuals. Linear relations are adjusted
for three distinct wavelength bins from the same night in the
Fig.~\ref{fig:extinction-airmass}.  In this example the night is
slightly non-photometric and the grey attenuation has a RMS of the
distribution of the order of 4\% (\cf Fig.~\ref{fig:extinction-alphas}
for a distribution of the $\delta T_{i}(\hat{z}, t)$ in airmass and in
time). Finally, the covariance map of all the adjusted parameters of the
night is presented in Fig.~\ref{fig:cov-matrix}.

\begin{figure*}
  \centering
  \includegraphics[width=.7\textwidth]{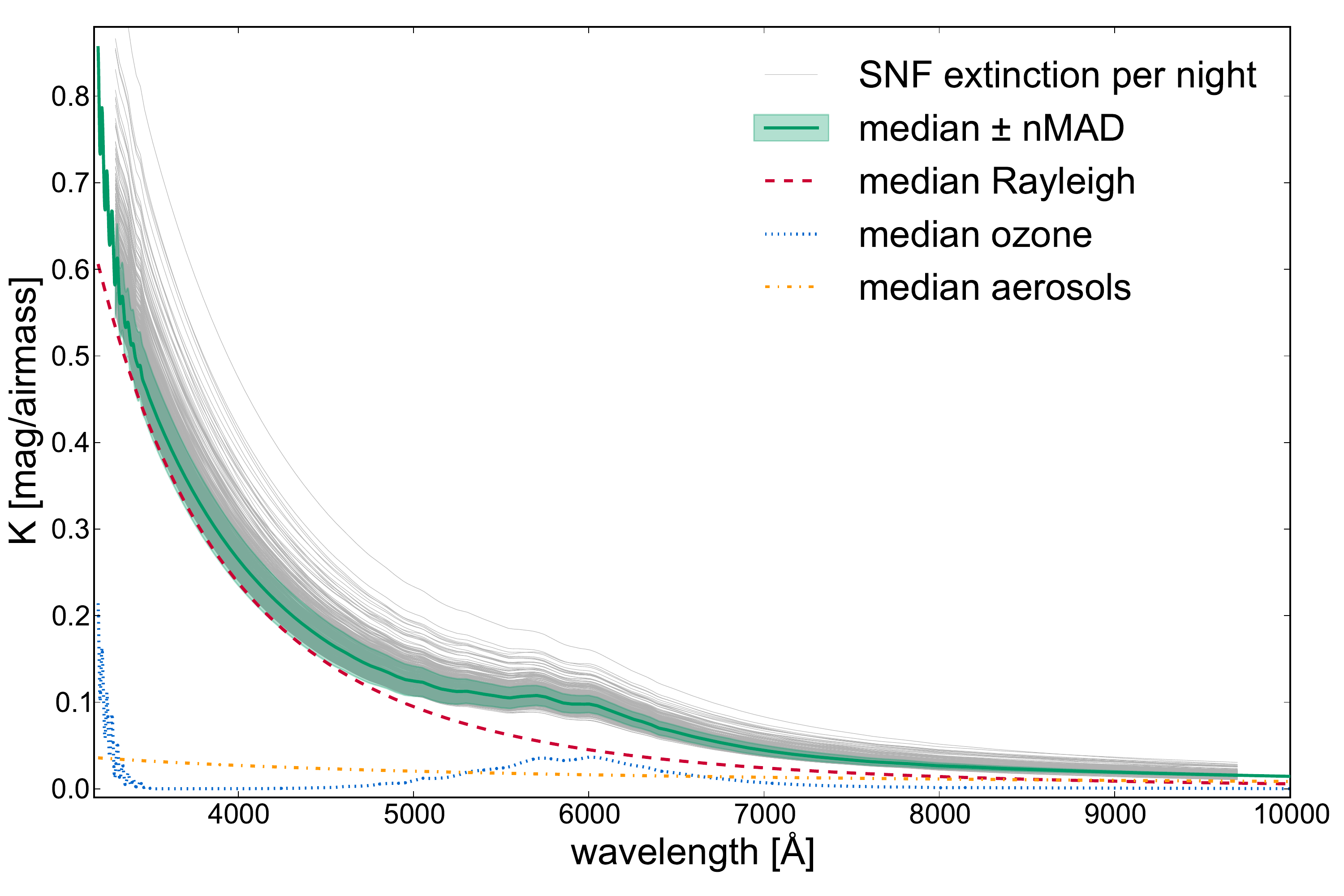}
  \caption{Typical extinction and its variability over the 7~years of
    the observation campaign. The superposition of all nightly
    extinction curves (grey) is shown, along with the median Mauna Kea
    extinction we derive (green).}
  \label{fig:extinction-variability}
\end{figure*}

\begin{figure*}
  \centering
  \includegraphics[width=0.9\textwidth]{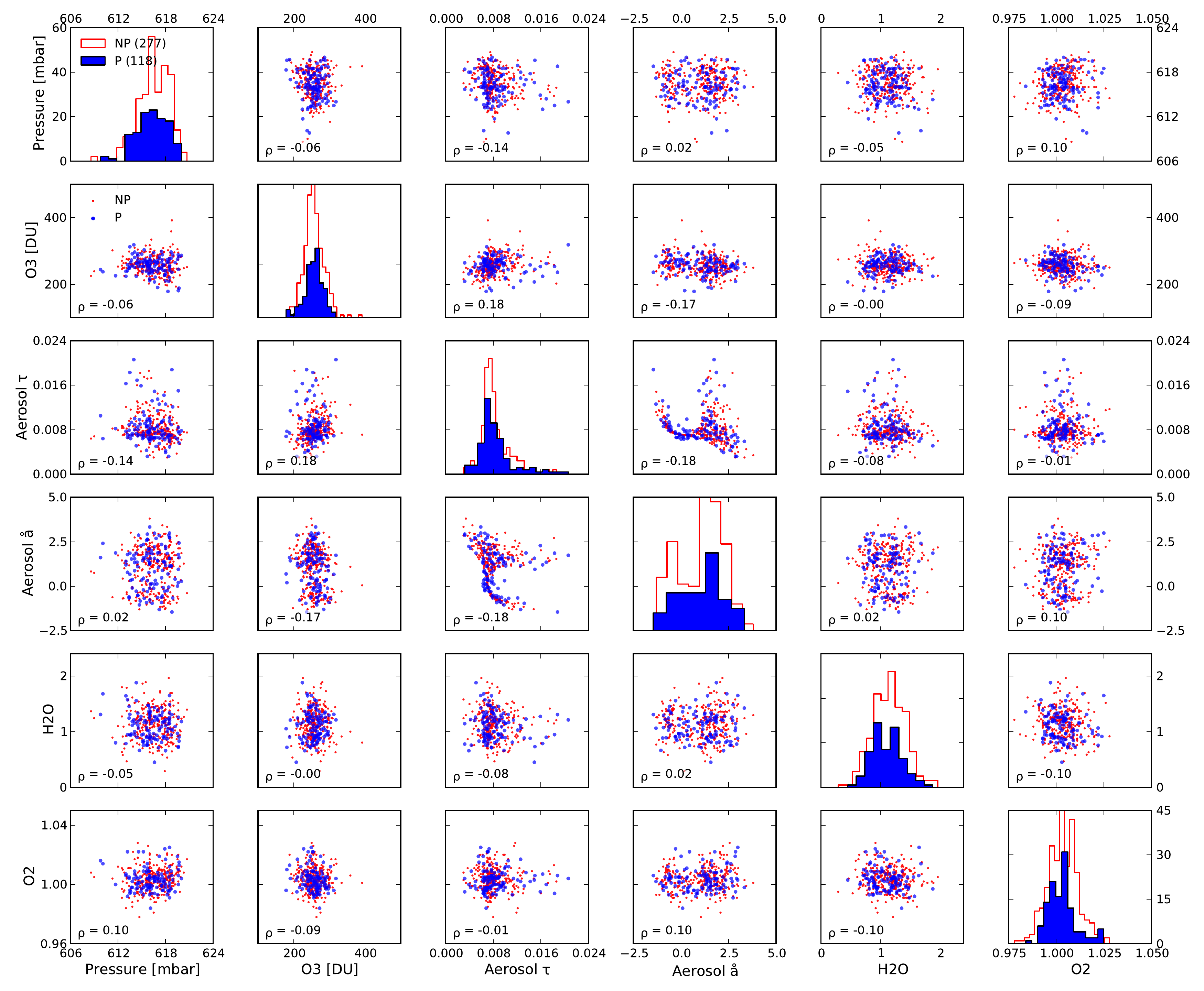}
  \caption{Correlations between extinction parameters for both
    photometric (blue circles) and non-photometric nights (red
    points). No strong correlations are seen between pressure, ozone, or
    telluric \HdO and \Od.  The aerosol parameters $\tau$ and $\aa$ are
    correlated because they represent the same physical component. The
    distributions of the parameter values (diagonal plots) also are
    similar on photometric and non-photometric nights. Note that the
    incidence of aerosol exponents near $\aa \sim 0$ is suppressed on
    non-photometric nights since the grey extinction term can completely
    compensate for this case.}
  \label{fig:ext-pars-correlation}
\end{figure*}

The 478~extinction curves computed using this method are presented in
Fig.~\ref{fig:extinction-variability}. Individual nights are plotted in
light grey whereas the median extinction, based on the median physical
parameters, is represented by the thick solid green line. The green band
represents the dispersion of the nightly extinction determinations.
Fig.~\ref{fig:ext-pars-correlation} shows the correlations
between all extinction parameters for both photometric and
non-photometric nights. No obvious correlations appear between the
parameters except for $\tau$ and $\aa$, which are
expected to be correlated since in combination they represent
the aerosol component of the extinction. This demonstrates the
independent behavior of the physical components.

\section{Accuracy and variability}
\label{sec:extinction-errors}

Having presented the fitting methodology and median results, we will now
discuss the accuracy of the results and what can be said about the
variability of the various physical components.

\subsection{Extinction accuracy}
\label{sec:computation-accuracy}

\begin{figure}
  \centering
  \includegraphics[width=\columnwidth]{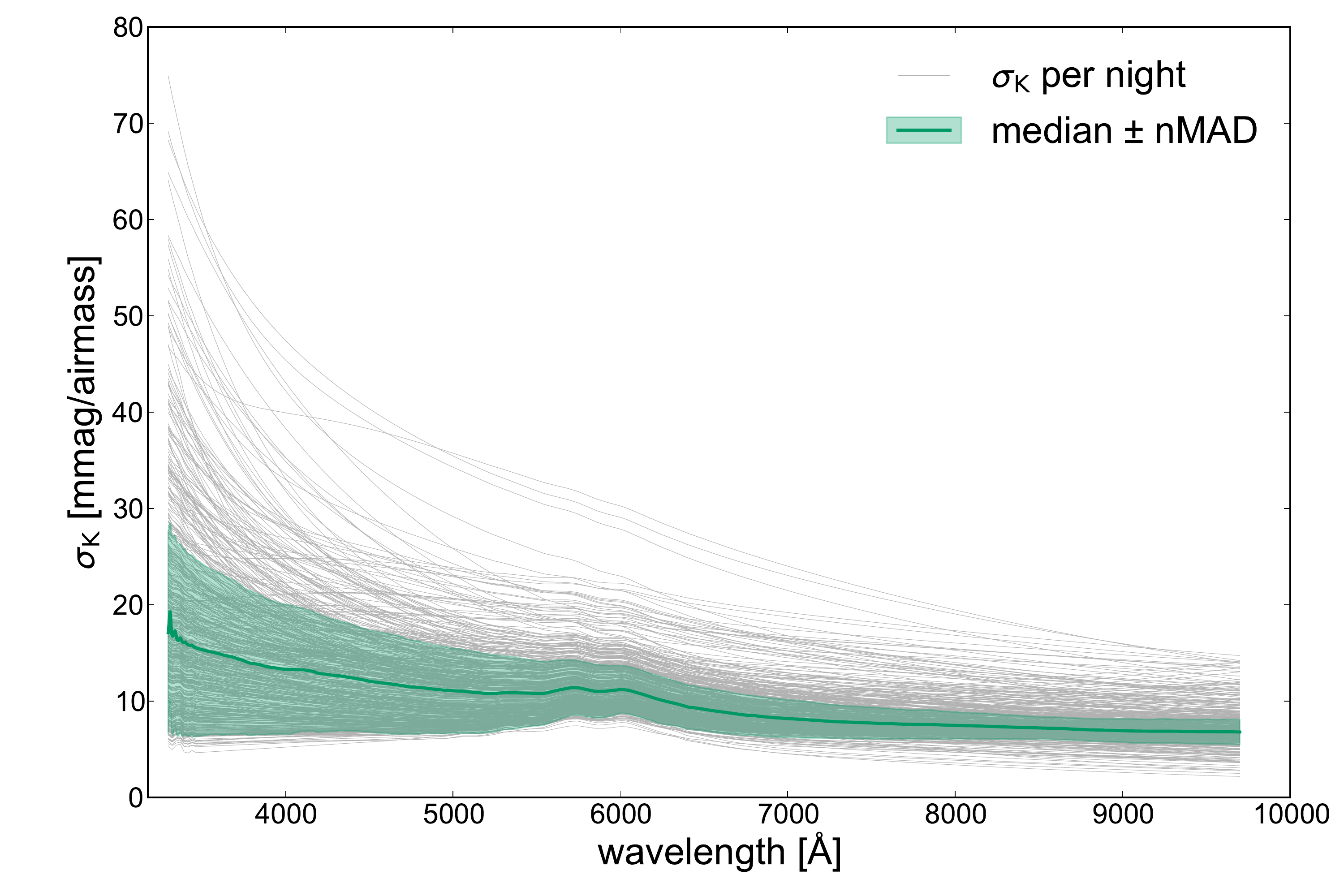}
  \caption{Error on the nightly extinction computations as a function of
    wavelength (light grey lines).  The mean error and RMS of the
    extinction curves distribution are shown in green.}
  \label{fig:extinction-errors}
\end{figure}

The 478~extinction error spectra are shown in
Fig.~\ref{fig:extinction-errors}. These curves represent our ability to
measure the extinction with the standard observations taken each night.
This is largely set by the numbers of standard stars and their airmass
range.  These curves do not represent natural variations in the
extinction, which are addressed below. The mean error, displayed as a
green line in the figure, decreases from its maximal value of
20~mmag/airmass in the UV to 7~mmag/airmass in the red.  This power law
behavior can be explained by the fact that the aerosol component is the
main source of variability allowed in our model.  A similar behavior is
observed for all the individual error spectra (light grey curves) with
various values of the power exponent. On nights with few standards or a
small airmass range, the errors are larger. On such nights, use of the
median extinction curve rather than the nightly curve should provide
better calibration.

\subsection{Rayleigh variability}
\label{sec:rayleigh-variability}

The average Rayleigh extinction component is shown in
Fig.~\ref{fig:RayleighVariability}.  During the course of the
observations the surface pressure at the summit of Mauna Kea showed a
peak-to-peak variation of 6~mbar around the average 616~mbar value.  The
nMAD of the variation was 2~mbar, which corresponds to an extinction
variation of 2~mmag/airmass at the blue edge of our spectral range.  The
peak-to-peak extinction variation still would be only 6~mmag/airmass.
These variations are negligible with respect to the aerosol or ozone
components and have no impact on the global extinction variability at
our required level of accuracy.

\subsection{Aerosol variability}
\label{sec:aerosol-variability}

The median aerosol component we derive is shown in
Fig.~\ref{fig:AerosolVariability}, superimposed on our aerosol
measurements from each night. Fortunately the median value is
very low, ranging from $\sim$40 down to $\sim$10~mmag/airmass over the
\snifs wavelength range.  Even so, aerosol variations are the largest
contributor to the variability of the extinction continuum from one
night to another, and these variations are also shown in
Fig.~\ref{fig:AerosolVariability}. The combined fluctuations in $\tau$
and the \ang exponent are responsible for variations that reach an
extreme of 0.4~mag/airmass at 3300~\AA{}. It should be noted that \snifs
is not operated when winds exceed 20~m/s, so there may be more extreme
aerosol excursions that our data do not sample.

\subsection{Ozone variability}
\label{sec:ozone-variability}

\begin{figure}
  \centering
  \subfloat{%
    \label{fig:RayleighVariability}%
    \includegraphics[width=\columnwidth]{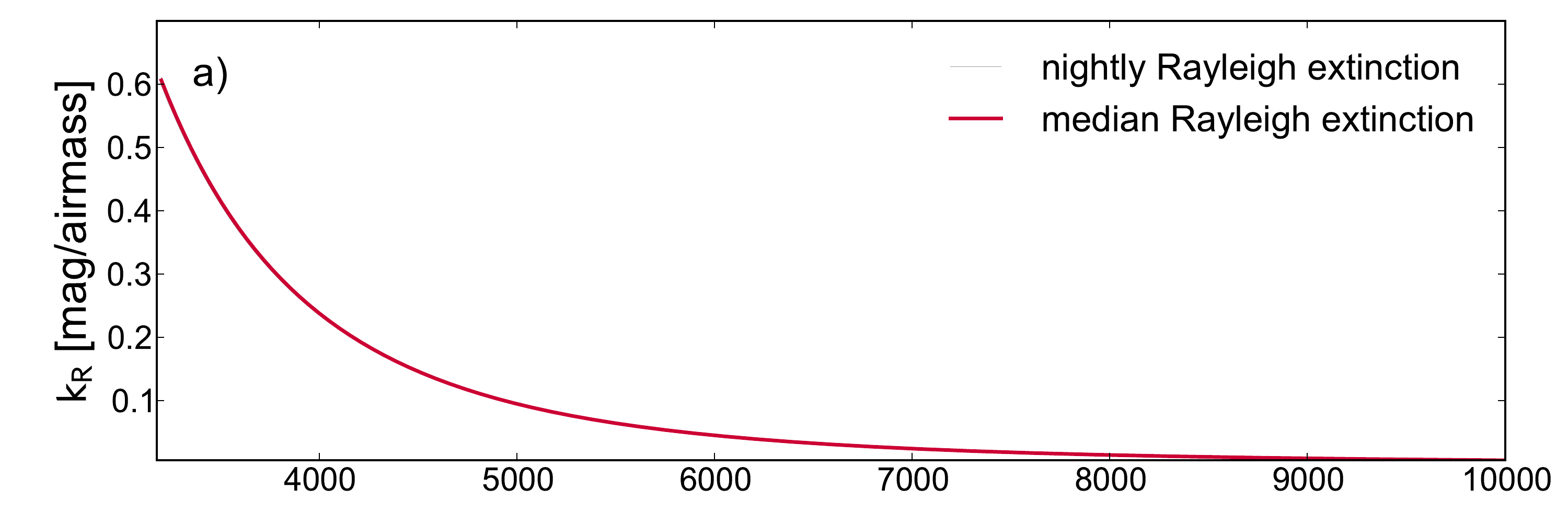}}\\
  \hspace{0mm}%
  \subfloat{%
    \label{fig:AerosolVariability}%
    \includegraphics[width=\columnwidth]{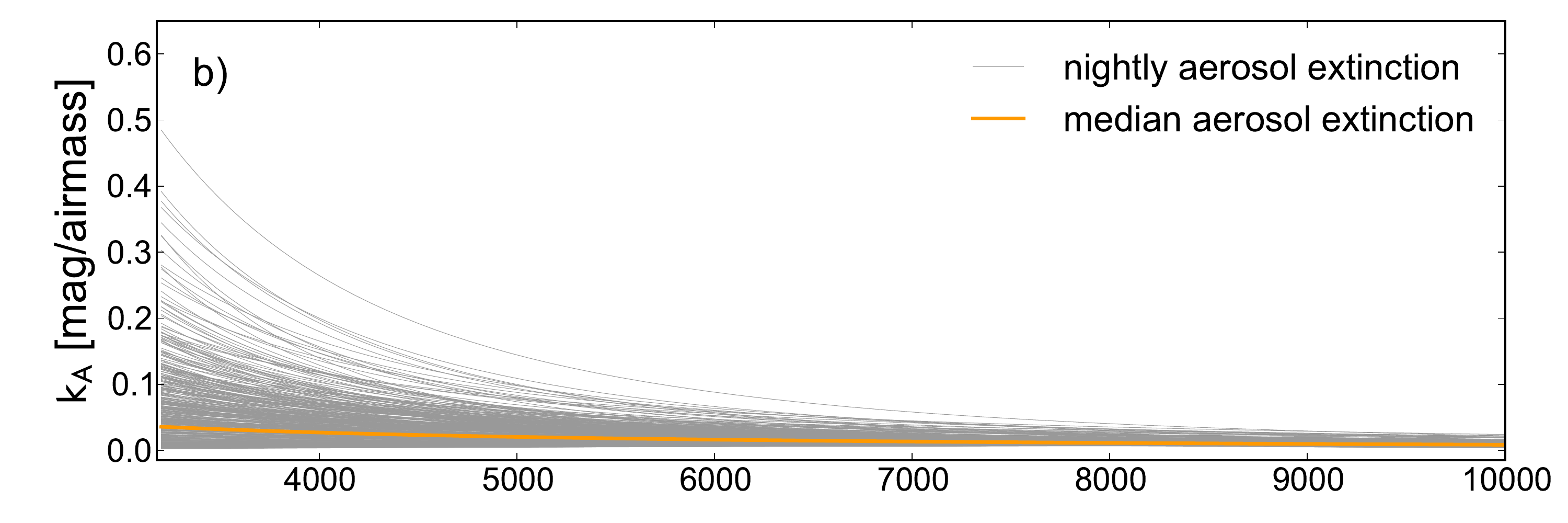}}\\
  \hspace{0mm}%
  \subfloat{%
    \label{fig:OzoneVariability}%
    \includegraphics[width=\columnwidth]{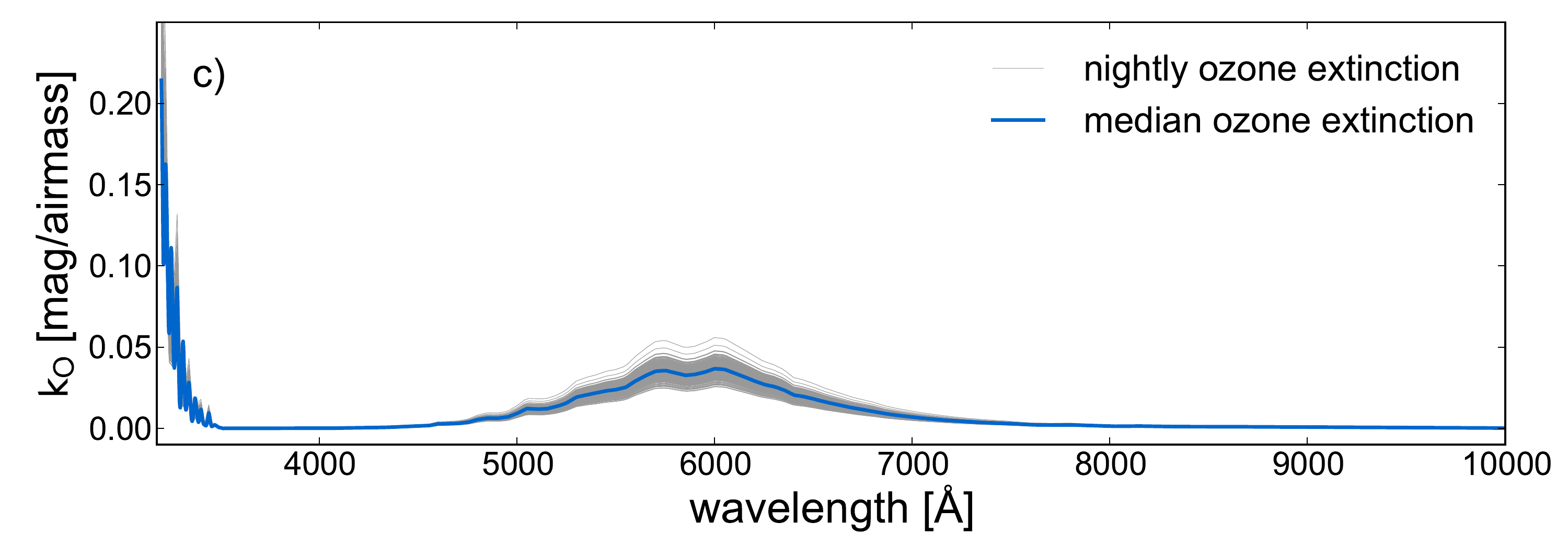}}%
  \caption{Nightly contributions of each physical component of the
    continuum extinction, showing their mean value and variability
    during the course of our observing campaign. All the nightly
    Rayleigh extinction curves (in grey) are within the width of the
    red median curve.}
  \label{fig:components-variability}
\end{figure}

The mean ozone component we derive is shown in
Fig.~\ref{fig:OzoneVariability}, superimposed on our ozone measurements
from each night. Over the Big Island the mean is 260~DU, a value lower than
the world-wide mean. The maximal peak to peak ozone variability in one
year can reach 60 to 80~DU. Much of this variation is scatter around a
clear seasonal trend attributable to the Quasi-Biennial Oscillation
(QBO), which has a mean amplitude around 20~DU.  Much of the remaining
variation is due to tropospheric winds \citep{steinbrecht03}. 20~DU is
only a 8\% variation, corresponding to 4~mmag/airmass for the ozone peak
at 6000~\AA{}.  We will show in \S~\ref{sec:ozone-mauna-loa}, this level
of variation currently is very difficult for us to detect. While this
level of uncertainty is unimportant for our science program, we could
better constrain ozone by using the \snifs standard star signal
available below 3200~\AA{}, where the Hartley \& Huggins band is much
stronger.

\subsection{Telluric line variability}
\label{sec:telluric-variability}

\begin{figure}
  \centering
  \includegraphics[width=\columnwidth]{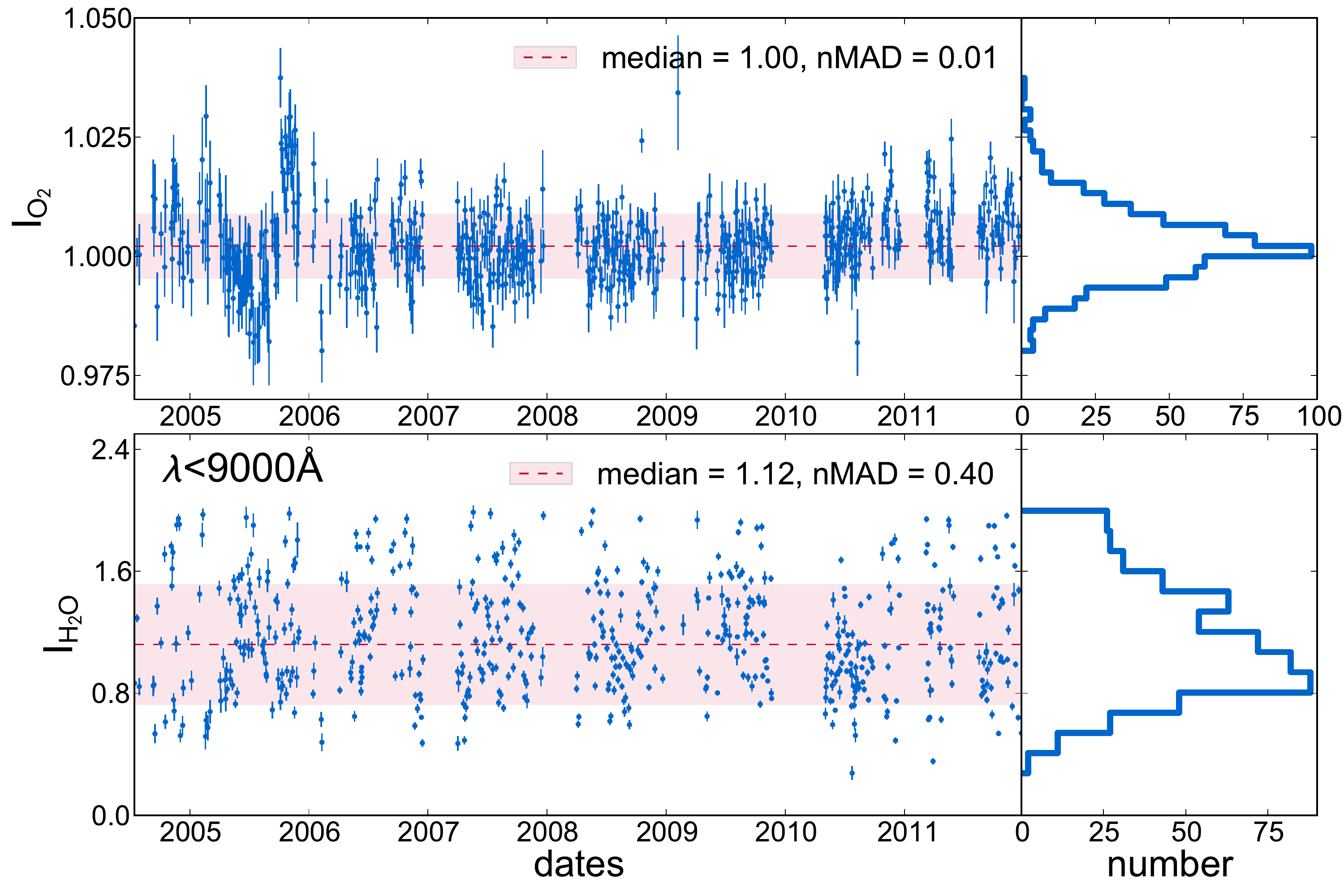}
  \caption{Distribution of the telluric strength over the \snf data set for
    each group of features. The strength represents the multiplicative
    factor needed to scale the template to the observed telluric features, and
    as such has no units.}
  \label{fig:telluric-intensity}
\end{figure}

The strength of the \Od and \HdO telluric features is displayed in
Fig.~\ref{fig:telluric-intensity}.  The fluctuations of the strength of
the \Od lines are remarkably small. This is fortunate given the strength
of the \Od A and B bands. On the other hand the strengths of the \HdO
lines vary widely. For some nights our current practice of assuming a
fixed \HdO strength for the entire night may be too simplistic. However,
Mauna Kea has less Precipitable Water Vapor than most ground-based
astronomical sites, and so far we have not encountered any serious
problems with this approximation. We explore this question further in
\S~\ref{sec:telluric-comparison}.

\begin{figure*}
  \centering
  \includegraphics[width=\textwidth]{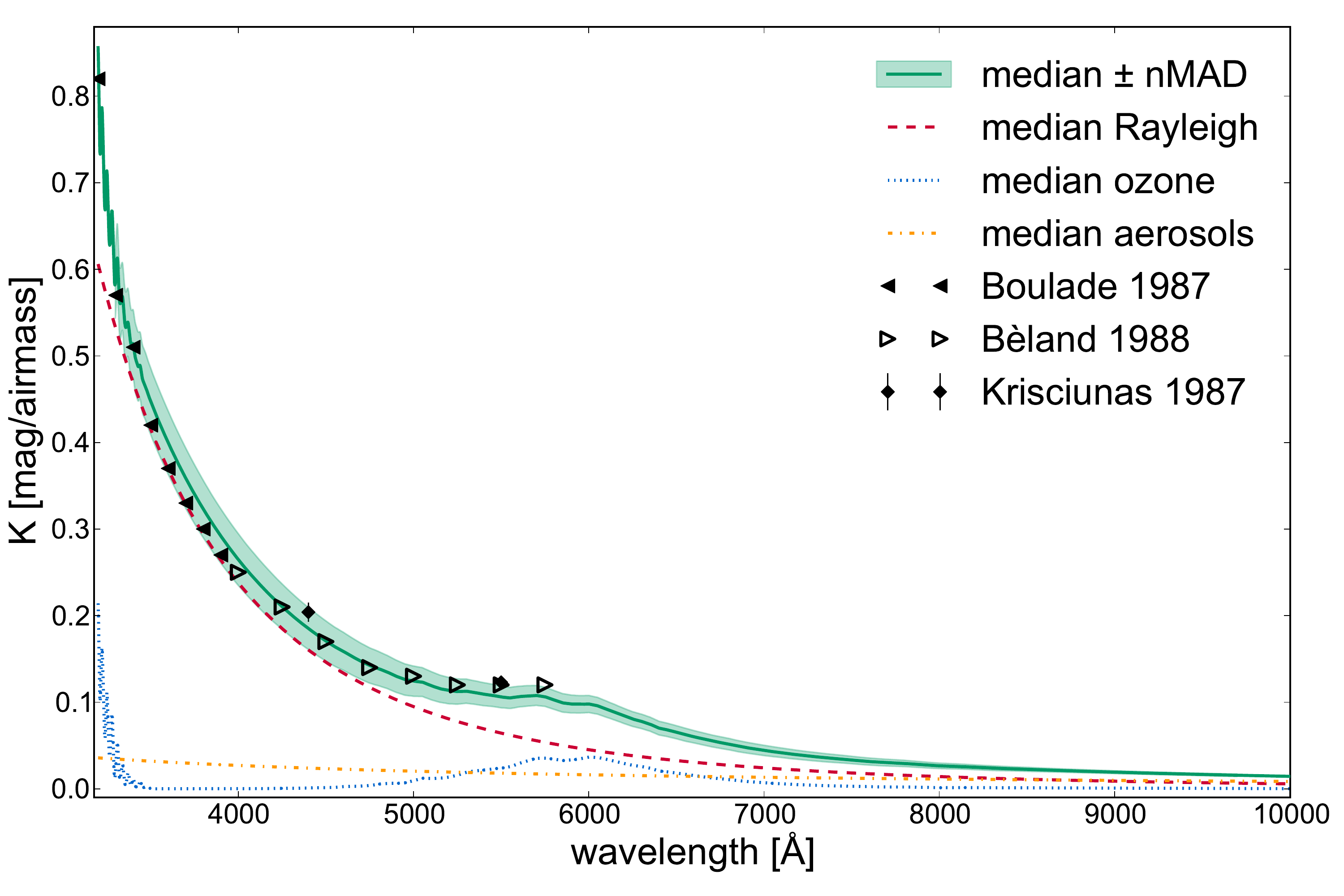}
  \caption{Mean \snf atmospheric extinction (solid line) and its
    physical components (dashed lines). For comparison we overplot the
    previous Mauna Kea extinction measures derived by \cite{Boulade87}
    (diamonds), \cite{Beland88} (triangles) and \cite{Krisciunas87}
    (stars).}
  \label{fig:extinction-comparison}
\end{figure*}

\subsection{Short time variability}
\label{sec:short-variability}

Due to our limited temporal sampling during each night, and the 2--3\%
achromatic scatter, it is difficult for us to detect extinction
variability of less than 1\% over the course of a night. As noted above,
Rayleigh scattering variations are negligible, and ozone variations are
nearly negligible with most of the variation occurring on seasonal
timescales. Aerosol variations thus remain the primary concern for
extinction variations during the night.

It should be noted that the $\delta T(\hat{z}, t)$ term, used on
photometric and non-photometric nights alike, is degenerate with aerosol
extinction having an \ang exponent of zero. Therefore, part of any
temporal aerosol variation will be absorbed.

The GONG site survey of Mauna Kea \citep{Hill94a, Hill94b} found
exceptional extinction stability. \cite{Mann11} have demonstrated
stability at the milli-magnitude level over the course of hours using
\snifs. Unfortunately, our extraction error is large enough (2 to 3\%)
to dominate our measurement of aerosol extinction variability (see
Fig.~\ref{fig:comp-extinction} for typical errors). We have nevertheless
tried to compare the mean aerosol extinction of time domains distributed
at the beginning and the end of the night, but we have no
evidence/indication whatsoever for a strong aerosol variation during the
course of the night. Finally, the various metrics we used in
\S~\ref{sec:photometricity} to determine the photometric quality of each
night are sensitive enough to alert us to transmission changes greater
than several percent. If we are able to reduce the achromatic standard
star extraction uncertainty, our sensitivity to temporal variations
would also improve.

Turning to other sites, for Cerro Tololo in Chile \cite{Burke10} showed
that aerosol variations were rather small during a night. \cite{Burki95}
showed that for the La Silla site in Chile the total $U$-band extinction
is correlated over a period of several days and that over one night the
auto-correlation drops by only 5\%. This implies a typical $\sim 2$\%
variation in extinction between the beginning and the end of the night.
In our formalism the $\delta T(\hat{z}, t)$ term can be used to test for
such temporal trends on photometric nights. These sites have similar
altitudes (2.1~km and 2.3~km, respectively), with inversion layers that
are near \citep{Burki95, Gallardo00} their summits. Since more aerosol
variability might be expected for these sites than for Mauna Kea, this
further justifies our assumption that nightly variability of
aerosols is relatively unimportant for Mauna Kea.

\section{Comparison to external measurements}
\label{sec:comp-meteo}

As noted in the introduction, there have been previous measurements of
the optical extinction above Mauna Kea, and we begin this section with a
comparison of our results to those previous
studies. Fig.~\ref{fig:extinction-comparison} shows our median Mauna Kea
extinction curve along with the observed fluctuations. Previous spectroscopic
determinations, which covered only the blue half of the optical window, are
overplotted as diamonds \citep{Boulade87} and stars
\citep{Beland88}. The broadband filter extinction measurements from
\cite{Krisciunas87} are also shown (triangles). These external sources,
taken more than 20~years prior, show excellent agreement with our own
Mauna Kea extinction curve.

For further comparison we turn to atmospheric sciences data from Mauna
Loa Observatory\footnote{\url{http://www.esrl.noaa.gov/gmd/obop/mlo/}}
\citep[MLO,][]{Price59, Price63}, operated by the U.S. network
\emph{Earth System Research Laboratory} of the National Oceanic and
Atmospheric Administration (NOAA).  MLO is situated on the side of Mauna
Loa mountain, 41~km south of Mauna Kea, at an altitude of 3400~m and
770~m below the summit.  The observatory possesses specific
instrumentation for aerosol and ozone measurements and we are
particularly interested in comparing these measurements of ozone
strength, aerosol optical depth and \ang exponent with our own
measurements obtained during nighttime on Mauna Kea.

\subsection{Ozone comparison}
\label{sec:ozone-mauna-loa}

\begin{figure}[!h]
  \centering
  \includegraphics[width=\columnwidth]{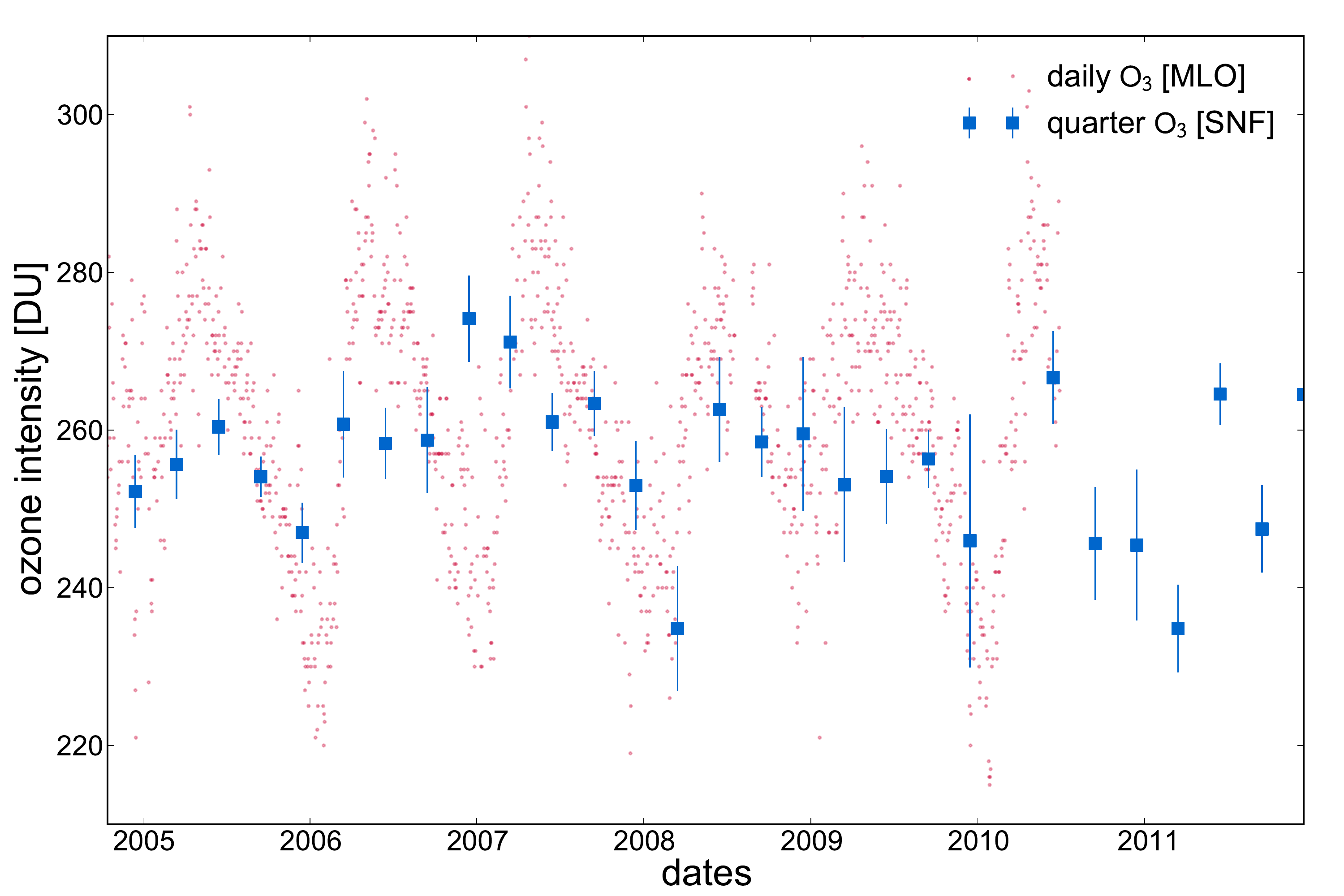}
  \caption{Total Column Ozone [Dobson units] from May 2004
    to December 2011 for the \snf quarterly weighted averages (blue
    squares) and the Mauna Loa Observatory daily measurements (red
    points). The error bars on the \snf points represent the standard
    errors on the quarterly weighted averages.}
  \label{fig:ozone-intensity}
\end{figure}

The Total Column Ozone (TCO) in the region (\ie the amount of ozone in
a column above the site from the surface to the edge of the
atmosphere) is measured at the
MLO\footnote{\url{http://www.esrl.noaa.gov/gmd/ozwv/dobson/mlo.html}}
three times per day during week days using a Dobson
spectro-photometer \citep{Komhyr97}.  The TCO comparison
between Mauna Loa and \snf data is presented in
Fig.~\ref{fig:ozone-intensity}.

The seasonal \Oz variability is well established by the Mauna Loa data,
with clear peaks corresponding to summer. The seasonal pattern observed
in the \snf ozone variability is less obvious. This is due to our use of
a prior. The typical nightly measurement uncertainty on $I_{\Oz}$ is
around 60~DU without the prior. Given our prior with standard deviation
50~DU (see Table~\ref{tab:priors}), the fit will return a value of
$I_{\Oz}$ roughly midway between the true value and the mean of the
prior (260~DU). This bias suppresses the amplitude we measure for the
seasonal variation in ozone, but in the worst case leads to a
spectrophotometric bias below 0.3\%. If desired, this bias could be
further minimized by using the Mauna Loa values as the mean for the prior on each
night, or by first estimating a prior on a quarterly basis from our own
data using an initial run without the prior, or by measuring \Oz using
SNIFS coverage of the Hartley \& Huggins band below 3200~\AA.

Overall, the behavior of our ozone measurements averaged over a quarter
seems consistent with the Mauna Loa measurements although being not as
accurate.

\subsection{Aerosol comparison}
\label{sec:aerosol-mauna-loa}

\begin{figure}
  \centering
  \subfloat{%
    \label{fig:angstrom-exponent}%
    \includegraphics[width=\columnwidth]{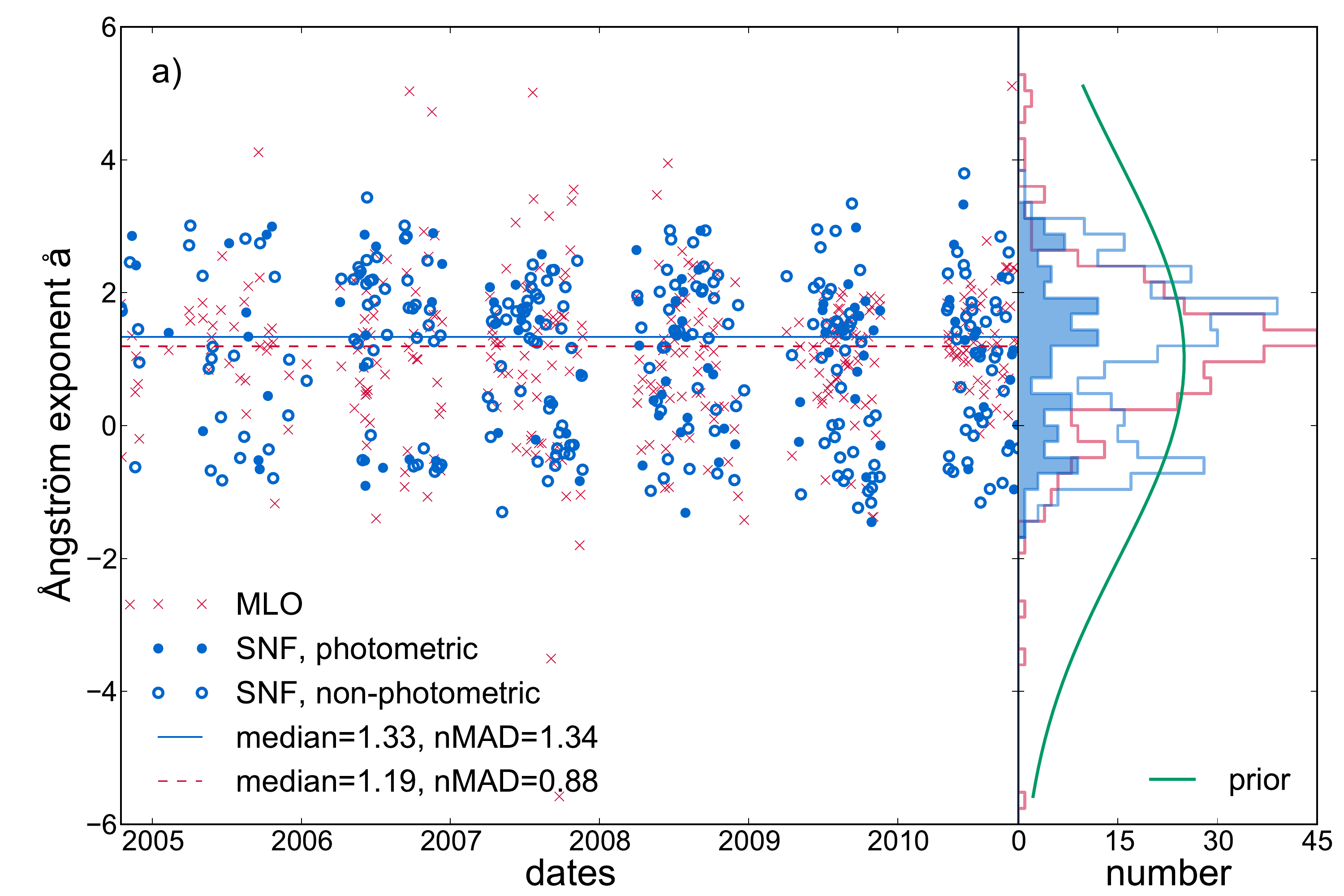}}\\
  \hspace{0mm}%
  \subfloat{%
    \label{fig:optical-depth}%
    \includegraphics[width=\columnwidth]{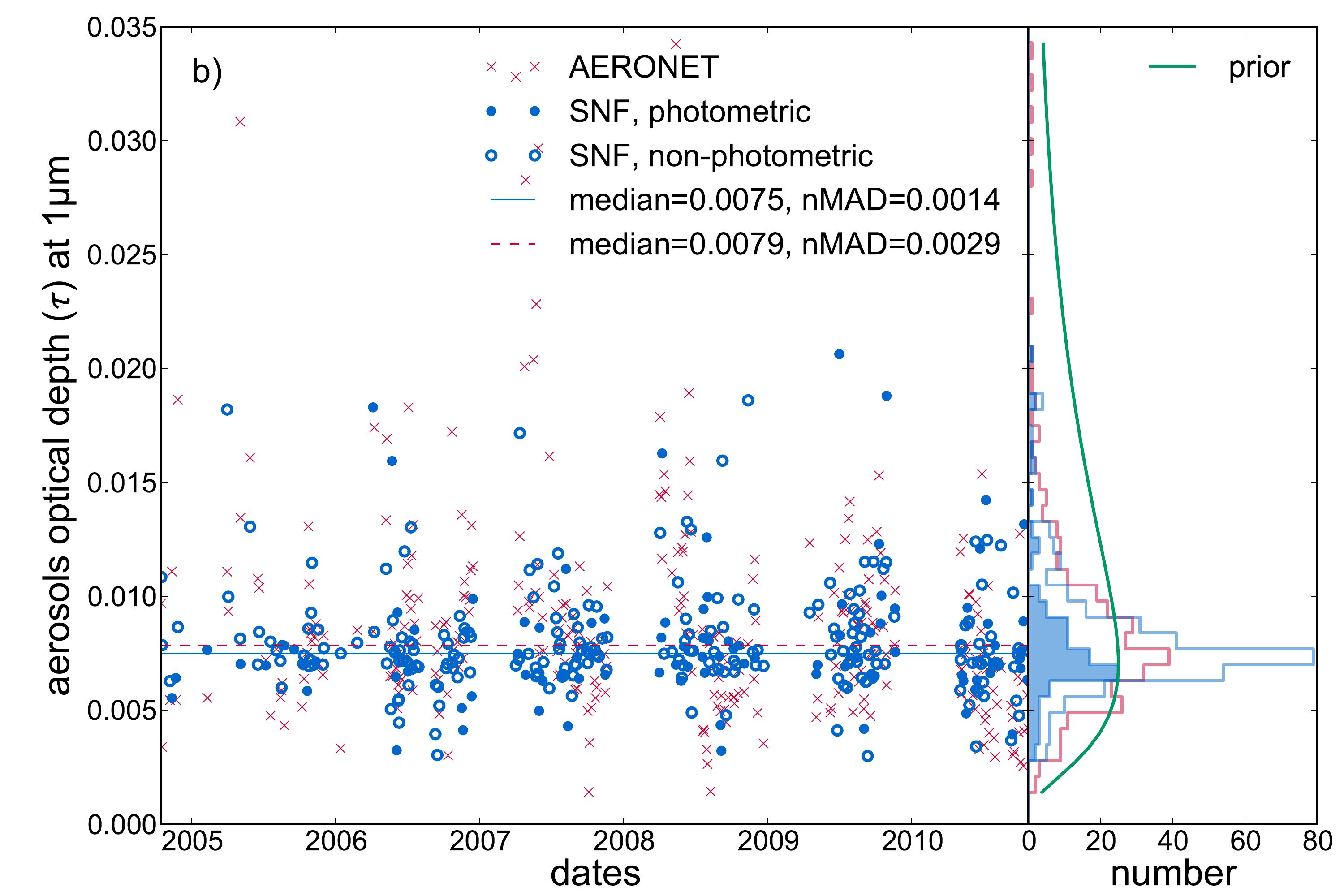}}%
  \caption{Aerosol \ang exponent
    \protect\subref{fig:angstrom-exponent} and optical depth
    \protect\subref{fig:optical-depth} distributions from May 2004 to
    December 2010 for the \snf (circles) and the Mauna Loa Observatory
    (crosses) measurements. Only the common nights of both data sets are
    presented. Mauna Loa aerosol data were not available for 2011 in
    time for our analysis.}
\end{figure}

The MLO ground-based aerosol data come from the \aeronet \citep[AErosol
RObotic NETwork.][]{Holben01, Smirnov01, Holben03}.  \aeronet uses wide
angular spectral measurements of solar and sky radiation measurements.
For these reasons, such aerosol measurements at Mauna Loa have only been
carried out during daytime, and usable data requires reasonably good
weather conditions (no clouds).  Figures~\ref{fig:angstrom-exponent} and
\ref{fig:optical-depth} show the comparisons between Mauna Loa data and
\snf data for the \ang exponent and the optical depth of the aerosol.

The \snf \ang exponent, $\aa$, from 2004 to 2011 is distributed between
$-2$ and 4. The fact that the Mauna Kea site is almost 1000~m higher
than MLO, and the fact that the \aeronet observations are carried out
during day time make a point to point comparison
difficult. Nevertheless, the general features of both distributions
still can be compared with each other: the mean \snf \ang exponent ($1.3
\pm 1.4$) is compatible with the one measured by \aeronet ($1.2 \pm
0.9$).

We see in Fig.~\ref{fig:optical-depth} that the \snf aerosol optical
depth mean value is slightly smaller than the Mauna Loa mean value
(still, they are compatible with each other), and the distribution does
not show the slight seasonal trend observed in the \aeronet data.
Nevertheless, we do expect to have more stable values since the
measurements are made at night far above the inversion layer, and the
mean optical depth value is compatible with aerosols from maritime
origins.

\subsection{Telluric absorption comparison}
\label{sec:telluric-comparison}

\begin{figure}
  \centering
  \includegraphics[width=\columnwidth]{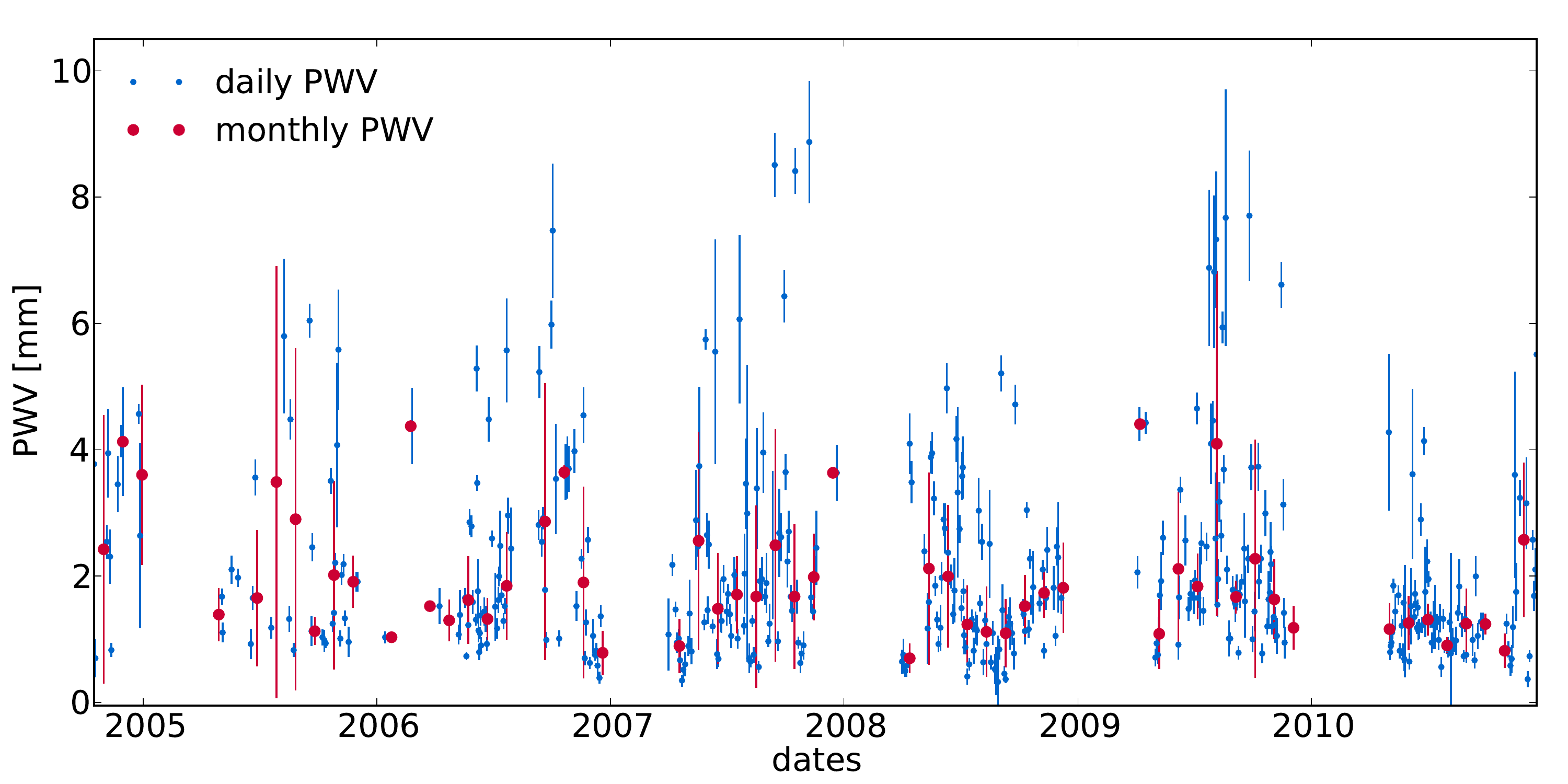}
  \caption{Precipitable Water Vapor (PWV) above the Mauna Kea summit
    during the \snf observation campaign.  The PWV is empirically
    computed \citep{Otarola10} from the atmospheric optical depth at
    225~GHz data by the 225~GHz tipping radiometer at the Caltech
    Submillimeter Observatory \citep{Peterson03}.}
  \label{fig:telluric-PWV}
\end{figure}

\begin{figure}
  \centering
  \includegraphics[width=\columnwidth]{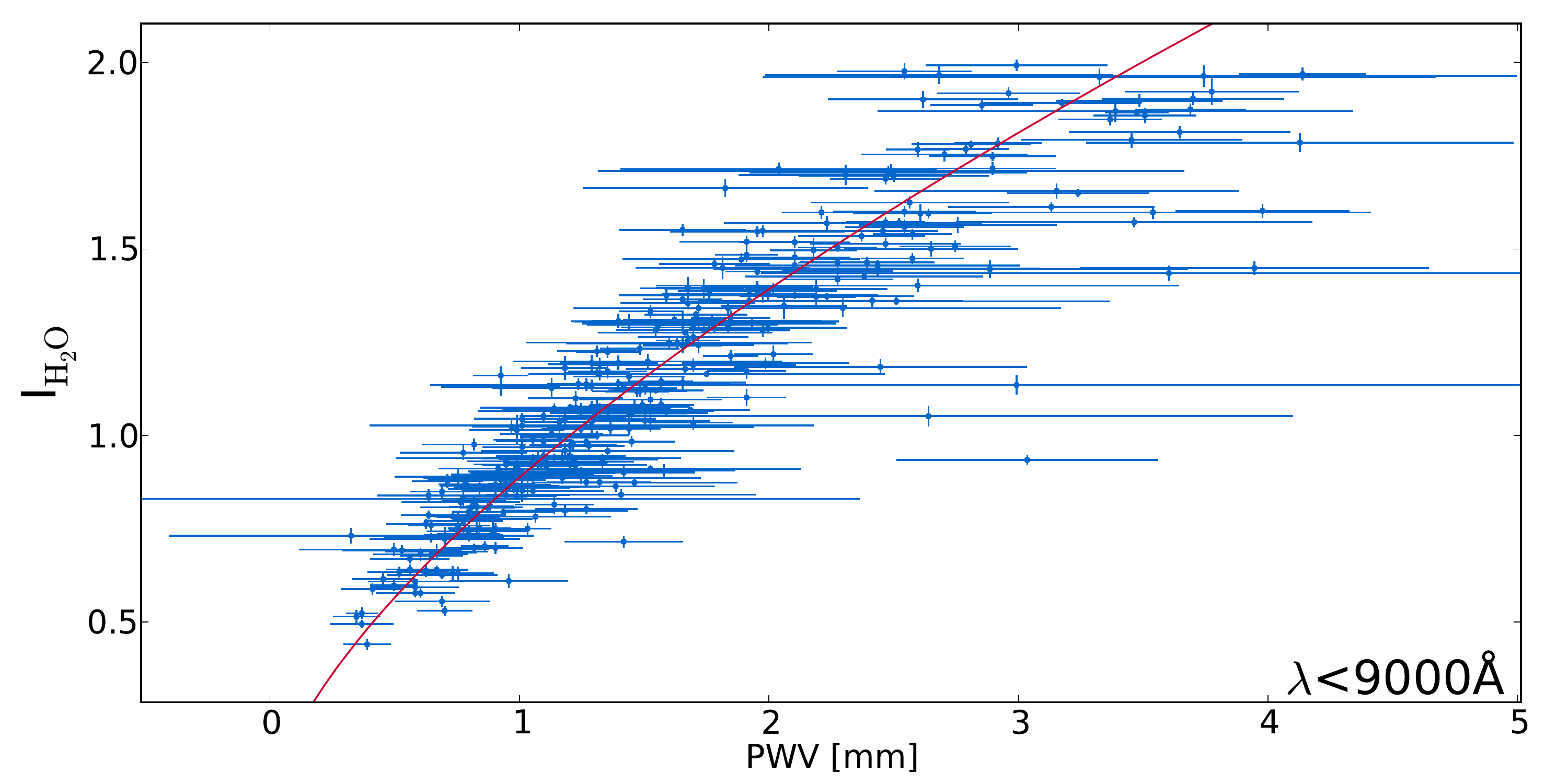}
  \caption{Correlation between the intensity $I_{\HdO}$ of the water
    telluric line computed from the \snf data and the Precipitable
    Water Vapor gathered from the Caltech Submillimeter
    Observatory. The red line represents the best power law fit of the
    distribution with an exponent value of 0.62 --- close to the water
    saturation parameter of 0.6 used in our procedure.}
  \label{fig:telluric-correlation}
\end{figure}

\cite{Patat11} showed that the water band equivalent width at
7200~\AA{} is well correlated with the Preciptable Water Vapor (PWV)
at the Paranal site in Chile.  In order to check the consistency of our
approach, we compare our water intensity measurement to the PWV amount
at Mauna Kea.  For that purpose, we used the optical depth at 225~GHz
data from the Caltech Submillimeter Observatory \citep[CSO,][]{Peterson03} and
the empirical formula from \cite{Otarola10} to compute the PWV at Mauna
Kea during the \snf observation campaign (\cf
Fig.~\ref{fig:telluric-PWV}). Plotting the computed telluric water
intensity (below 9000~\AA{}) with respect to the PWV (\cf
Fig.~\ref{fig:telluric-correlation}), we found that both quantities
were highly correlated, reinforcing the findings of \cite{Patat11} and
validating our water telluric correction approach. Furthermore, using an
orthogonal distance regression we find a saturation exponent of
$\rho=0.62\pm0.01$ from the best power law describing the distribution
(\cf red curve in Fig.~\ref{fig:telluric-correlation}). This is in
excellent agreement with the value $\rho_{\HdO}=0.6$ determined from the
airmass dependence of $I_{\HdO}$ in the \snf dataset.

\section{Revisiting standard assumptions}
\label{sec:discussion}

A major strength of our analysis comes from the homogeneity and temporal sampling of the \snf data
set. It consists of a large number of spectro-photometric standard star
spectra observed, processed and analyzed in a uniform way. In
addition, the sampling candence is frequent and covers a
long period of time. Observations were taken under a wide range of atmospheric conditions,
including drastically non-photometric nights, which allows us to test
many of the standard assumptions used in spectrophotometry, some of
which we employed in \S~{\ref{sec:formalism}}. In particular, the range
of conditions allows us to check our atmospheric model under conditions
of strong extinction by clouds, something that has not been done before
with spectroscopy simultaneously covering the full ground-based optical
window.

\subsection{Grey extinction}
\label{sec:discuss-grey}

\begin{figure}
  \centering
  \includegraphics[width=\columnwidth]{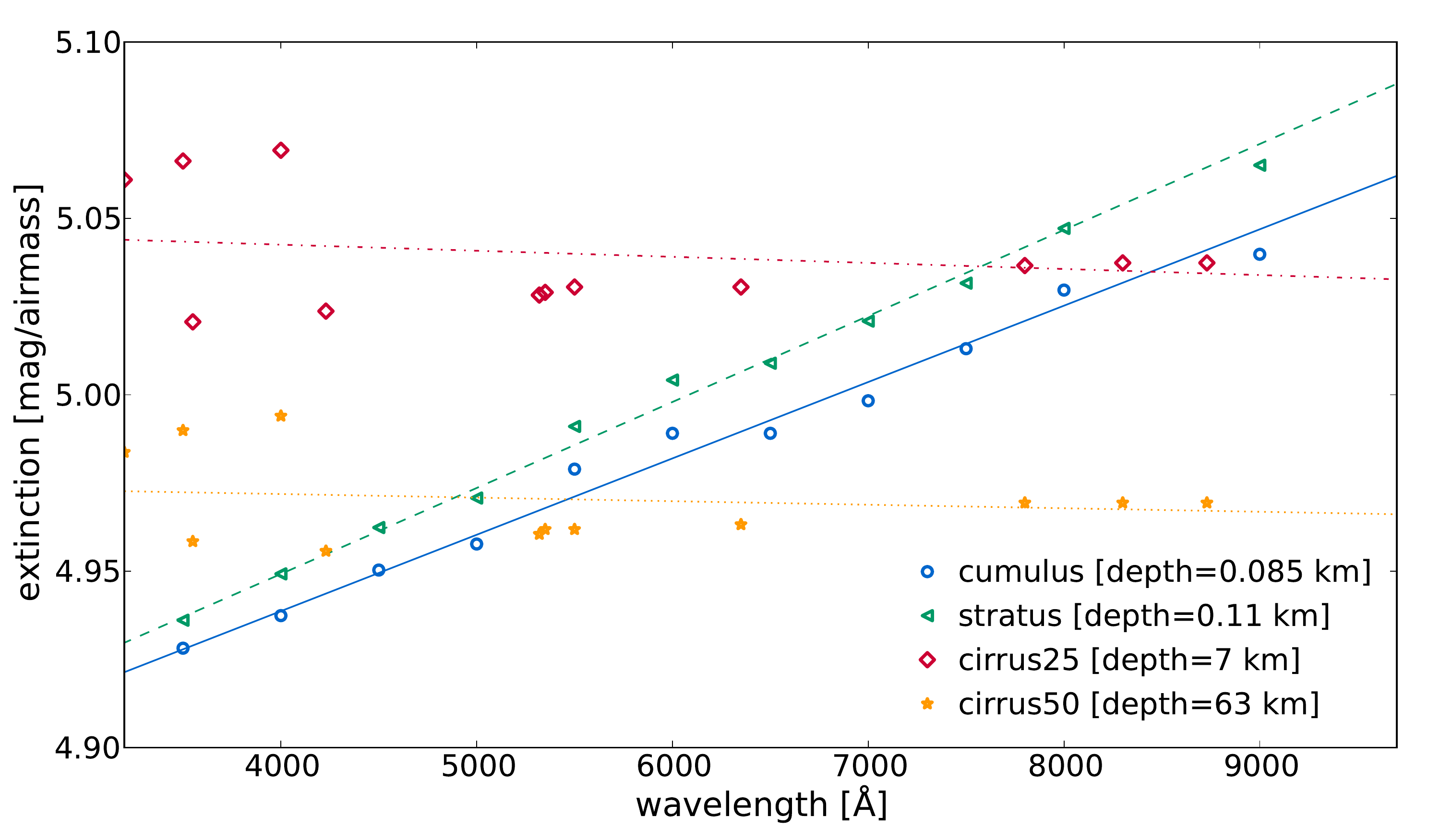}
  \caption{Simulations using OPAC of the wavelength dependence for
    typical clouds (in maritime environment) through very strong ---
    $\sim5$ magnitudes --- of extinction.  Even with such strong
    extinction, the trend from 3200~\AA{} to 9700~\AA{} is of the order
    of 3\% for cumulus and stratus and negligible for cirrus.}
  \label{fig:clouds-opac}
\end{figure}

Using the software package \emph{Optical Properties of Aerosols and
  Clouds} \citep[OPAC,][]{Hess98}, we simulated $\sim5$~mag of
extinction by clouds. While useful observations under such conditions
would be unlikely, this large opacity makes it possible to detect any
non-grey aspect of cloud extinction.  We analyzed three types of cloud
(cumulus, stratus and cirrus) in a maritime environment using the
standard characteristics from OPAC. The results, as shown in
Fig.~\ref{fig:clouds-opac}, demonstrate that extinction from cumulus
and stratus do exhibit a small trend in wavelength. But even for such
strong extinction of 5~mag/airmass the change in transmission from
3200~\AA{} to 10000~\AA{} is below 3\%. Cirrus with less than 1~mag
of extinction is the most common cloud environment that still allows
useful observing, and Fig.~\ref{fig:clouds-opac} demonstrates that
such clouds would be grey to much better than 1\%.

\begin{figure}
  \centering
  \includegraphics[width=\columnwidth]{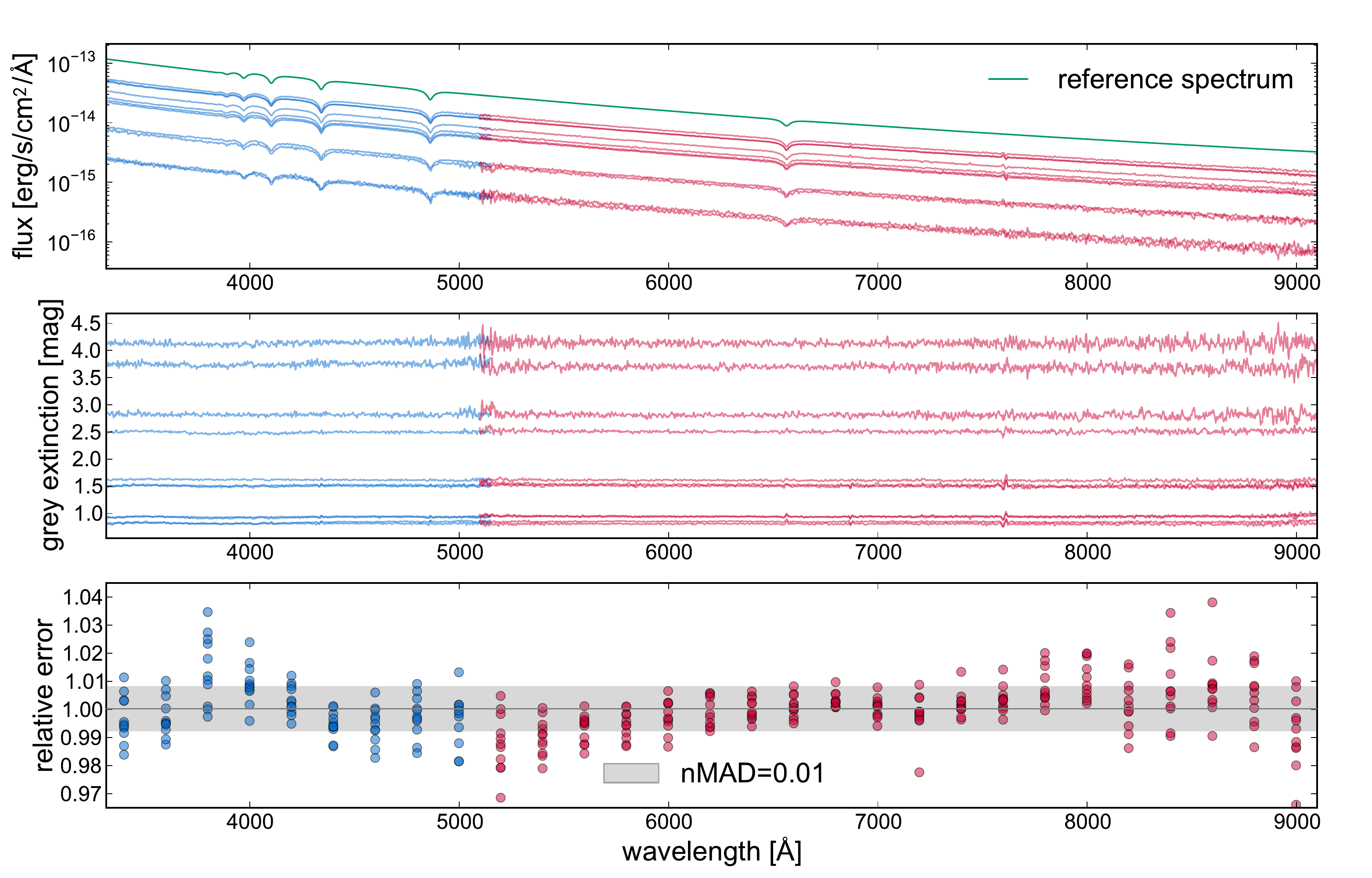}
  \caption{\snf observations of the standard star GD71 under various
    level of cloud extinction during non photometric nights.  The top
    panel shows the flux-calibrated spectra from both spectroscopy
    channels, without grey extinction correction, compared with the
    standard star reference template (green).  The middle panel shows
    the additional extinction $\delta T(\lambda, \hat{z}, t)$.  The
    bottom panel shows the variation with wavelength of the ratios
    spectra/reference. It is notable that even with strong extinction by
    clouds the transmission is compatible with grey extinction ($\delta
    T(\lambda, \hat{z}, t) = \delta T(\hat{z}, t)$) at the $\sim1$\%
    level (grey band).}
  \label{fig:clouds-extinction}
\end{figure}

Although atmospheric conditions such as these are usually avoided or
inappropriate for observations, we wanted to check whether or not such
an effect was visible in our data. Therefore, we gathered
11~observations (\cf Fig.~\ref{fig:clouds-extinction}) of the standard
star GD71 affected by different levels of cloud extinction (from 0.7
to 4.5~mag/airmass).  We calibrated these spectra without introducing
our grey extinction correction factor, $\delta T$. We then performed a
$\chi^{2}$ test that showed that the resulting extinction curves are
compatible with a constant to better than 1\% across the full
wavelength coverage of \snifs.

Our findings using spectroscopy agree with theoretical expectations
and the OPAC clouds models.  They also agree with the observational
analysis by \cite{Ivezic07} based on repeated broadband filter images
of the same \emph{Sloan Digital Sky Survey} (SDSS) field, which show
only a very weak dependence of the photometric residuals with color,
even through several magnitudes of extinction by clouds, and even when
comparing the flux in the $U$ and $Z$ bands. Similarly \cite{Burke10}
find little sign of color variation from $g-r$, or separately $r-i$,
$i-z$ and $z-y$, colors synthesized from their spectroscopic
measurements.

\begin{figure}
  \centering
  \includegraphics[width=\columnwidth]{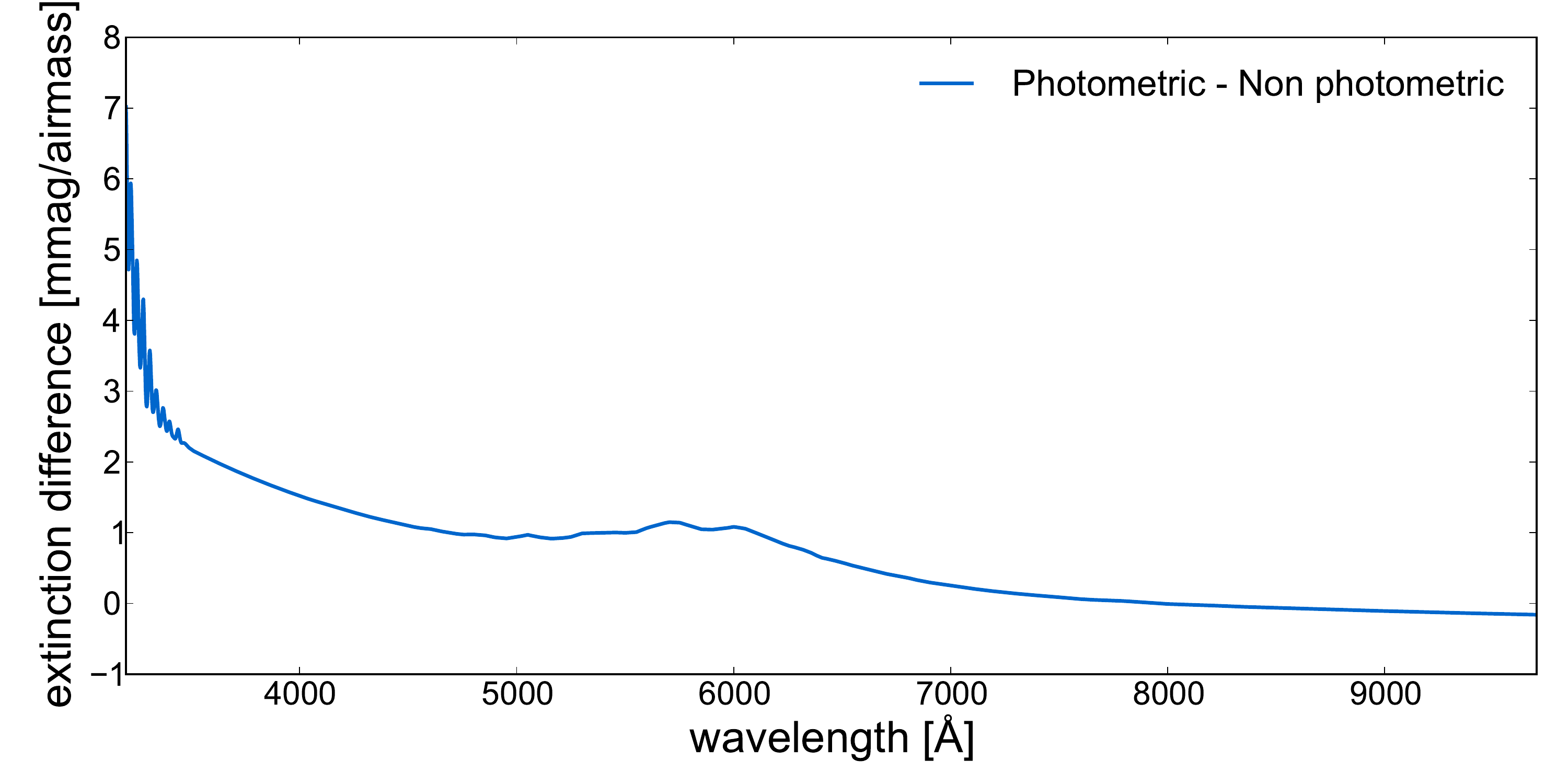}
  \caption{The difference, in [mag/airmass], between the median
    extinction measured on photometric nights and the median extinction
    measured on non-photometric nights.}
  \label{fig:P_vs_NP}
\end{figure}

As a final check, we can compare the median extinction curves from
photometric and non-photometric nights. Fig.~\ref{fig:P_vs_NP} shows
the difference of these two extinction curves. The agreement is much
better than 1\% over the full optical window. The smooth trend is due to
the aerosol component of our extinction model, and may be a hint either
of slight coloring due to clouds or a small difference in the aerosol
size distribution with and without clouds. A small residual due to ozone
is also apparent.

Altogether, we confirm that the treatment of cloud extinction being
grey remains valid at the 1\% level or better.

\subsection{Stability of the instrument}
\label{sec:stability-instrument}

In \S~\ref{sec:formalism} we assumed that the instrument calibration,
$C(\lambda, t)$, would be stable over the course of a night.  Here we
provide further justification for this assumption.

To begin, we note that there are several possible sources that could
lead to changes in the instrument calibration. Dust accumulation on
exposed optics alone can amount to 1~mmag/day per surface,
gradually reducing the reflectivity of the telescope mirror and
transmission of the spectrograph entrance window.  In our case, the
telescope tube is closed and the observatory dome is unvented, so dust
accumulation on the telescope optics is minimized. The telescope mirrors
are regularly cleaned with CO$_{2}$ ``snow'' and realuminized
occasionally.

Because the spectrograph optics are enclosed, their transmission is
expected to be stable (except for the dichroic beam-splitter which shows
changes with humidity --- an effect corrected in our pipeline).  The
quantum efficiency of the detectors will depend on the cold-head
temperature at the very reddest wavelengths, where phonons can assist an
electron into the conduction band. The electronics could experience
drift, though our gain tests using X-rays in the lab and photon transfer
curves \emph{in situ} do not show this.

To experimentally address this issue we have measured the stability of
$C(\lambda, t)$ over the course of several nights under photometric
conditions.  In this test $C(\lambda, t)$ was found to be consistent
from one night to another at better than the percent level using the
method developed in \S~\ref{sec:multi-stand-approach}.

\begin{figure}
  \centering
  \includegraphics[width=\columnwidth]{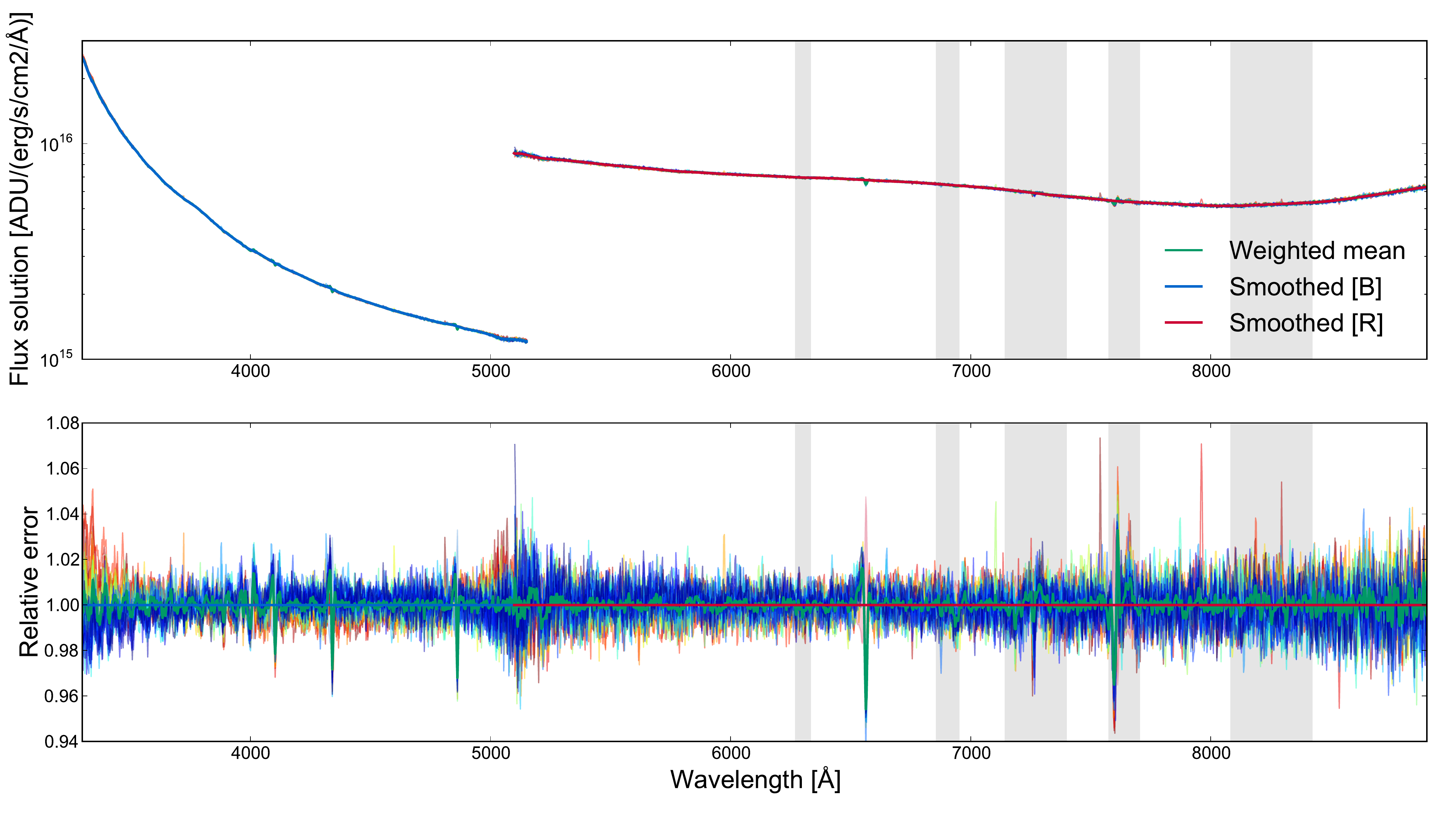}
  \caption{Instrument calibrations, $C(\lambda)$, for 23~observations of
    GD71 from the same highly non-photometric night (top
    panel). Relative error of each instrument calibration with respect
    to the mean value (bottom panel).  The mean RMS over the whole
    wavelength range is $~0.007$ and the features are due to spectral
    resolution issues between the spectra and their reference. Poorer
    accuracy due the water telluric band in the region redward
    8900~\AA{} is noticeable.}
  \label{fig:flux-solution-stability}
\end{figure}

In a separate test we compared $C(\lambda, t)$ using 23~standard stars
observed throughout the same highly non-photometric night.  When
divided by their mean, we find that the RMS of the distribution is
below the percent level.  In this case, because we allow for clouds,
only chromatic effects can be tested.
Fig.~\ref{fig:flux-solution-stability} shows the excellent agreement
throughout the night.

Of course it is important to keep in mind that major events, such as
cleaning or realuminization of the mirror, or instrument repairs, can
change the value of $C(\lambda, t)$ significantly. The desire to avoid
the need to explicitly account for such events, and the good nightly
consistency demonstrated above, justifies our choice to assume that
$C(\lambda, t) = C(\lambda)$ \emph{during} a night, while enabling it to change
from one night to another.

\subsection{Line of sight dependency}
\label{sec:line-sight}

One assumption that we have not directly tested is the dependence, or
lack thereof, of extinction on the viewing direction. Rayleigh and ozone
extinction variations across the sky will be negligible. Aerosols,
transported by winds, can be variable and so could vary across the
sky. Due to the very low level of aerosols typically encountered over
Mauna Kea, detection of such aerosol variations would require an
intensive campaign with \snifs, or better yet, with a LIDAR system. The
very low rates of change in aerosol optical depth found in the \aeronet
data for nearby Mauna Loa \citep[\eg][]{Stubbs07} suggest that
variations across the sky will be small or rare. Even for Cerro Tololo,
a site at significantly lower altitude and surrounded by dry mountains
(the Andes foothills) rather than ocean, \citet{Burke10} find little
evidence for aerosol gradients across the sky.  Thus, while we have not
been able to directly test the assumption of sightline independence for
our extinction measurements, any residual effects are likely to be
small. On non-photometric nights the achromatic portion of any sightline
dependence would be absorbed into the grey extinction term. In any
event, any such spatio-temporal variations will be averaged away across
our large dataset.

\section{Conclusions}
\label{sec:conclusions}

We have derived the first fully spectroscopic extinction curve for Mauna
Kea that spans the entire optical window and samples hundreds of
nights. The median parameters, tabulated in
Table~\ref{tab:final-params}, and the median curve, tabulated in
Table~\ref{tab:atmospheric_extinction}, can be used to correct past and
future Mauna Kea data for atmospheric extinction.  Comparison of our
median extinction curve shows good agreement in regions of overlap with
previously published measurements of Mauna Kea atmospheric extinction
\citep{Boulade87, Krisciunas87, Beland88} even though the measurements
are separated by roughly 20~years.

Our large dataset of 4285~spectra of standard stars collected over
7~years means that a wide range of conditions has been sampled. Using
this, we estimate the per-night variations on the extinction curve (see
Table~\ref{tab:atmospheric_extinction}), and this can be employed by
others to estimate the uncertainty in their data when using the median
extinction curve rather than deriving their own nightly flux
calibration. This is especially important for Mauna Kea due to the
presence of several of the world's largest telescopes, for which time
spent on calibration is often at odds with obtaining deeper
observations, and where calibration to a few percent is often deemed
adequate by necessity.

The method providing the extinction we have introduced in this paper has
several notable features. First, it borrows techniques long used in the
atmospheric sciences to decompose the extinction into its physical
components. This allows robust interpolation over wavelengths affected
by features in the spectra of the standard stars, and it accounts for
even weak extinction correlations spanning across wavelengths that would
otherwise be hard to detect. We have also utilized a method of dealing
with clouds by which the non-cloud component of the extinction still can
be measured. This enables chromatic flux calibration even through
clouds, and with an external estimate of the cloud component alone (such
as through the \snifs Multi-filter ratios) allows accurate flux
calibration even in non-photometric conditions.

Because of these desirable properties, we encourage the broader use of
this technique for flux calibration in astronomy. Furthermore, we note
that this parametric template is especially easy to use for simulating
the flux calibration requirements for future surveys (The atmospheric
extinction computation code is available from the \snf software web site
\url{http://snfactory.in2p3.fr/soft/atmosphericExtinction/}).

Finally, we have used our large homogeneous dataset to examine several
standard assumptions commonly made in astronomical flux calibration, like
the greyness of the clouds, and we confirmed our instrument
stability. These assumptions are usually difficult for most astronomers
to verify with their own data, which usually spans just a few nights. We
draw attention to the fact that the altitude of Mauna Kea and the
generally strong inversion layer sitting well below the summit helps
minimize the impact of aerosols and water vapor. These features are
often neglected relative to metrics related to image quality, but they
help make Mauna Kea an excellent site also for experiments requiring
accurate flux calibration.

\begin{table}
  \caption{Summary of the median values for the various adjusted
    physical quantities. For $\delta T$ which represents the cloud
    extinction, the absolute scale mean value is meaningless (as
    explained in \ref{sec:priors}), only variability could
    have an interest as an estimate of grey fluctuations during
    non-photometric conditions.}
  \label{tab:final-params}
  \centering
  \begin{tabular}{cc@{ $\pm$ }cl}
    \hline
    \hline
    Parameter & Median & nMAD & Description \\
    \hline
    $\delta T$ & 1.04 & 0.34 & grey extinction (mean \& RMS) \\
    $I_{\Oz}$ & 257.4 & 23.3 & \Oz intensity \\
    $\tau$ & 0.0084 & 0.0014 & aerosol optical depth \\
    $\aa$ & 1.26 & 1.33 & aerosol \ang exponent \\
    $\rho_{\Od}$ & 0.58 & 0.04 & telluric \Od saturation \\
    $\rho_{\HdO}$ & 0.60 & 0.27 & telluric \HdO saturation \\
    \hline
  \end{tabular}
\end{table}

%%%%%%%%%%%%%%%%%%%%%%%%%%%%%%%%%%%%%%%%%%%%%%%%%%%%%%%%%%%%%%%%%%%%%%

%%%%%%%%%%%%%%%%%%%
% Acknowledgments %
%%%%%%%%%%%%%%%%%%%

\begin{acknowledgements}
  We are grateful to the technical and scientific staff of the
  University of Hawaii 2.2-meter telescope for their assistance in
  obtaining these data. D.~Birchall assisted with acquisition of the
  data presented here. We also thank the people of Hawaii for access to
  Mauna Kea.  This work was supported in France by CNRS/IN2P3,
  CNRS/INSU, CNRS/PNC, and used the resources of the IN2P3 computer
  center; This work was supported by the DFG through TRR33 ``The Dark
  Universe'', and by National Natural Science Foundation of China (grant
  10903010). C. WU acknowledges support from the National Natural
  Science Foundation of China grant 10903010. This work was also
  supported in part by the Director, Office of Science, Office of High
  Energy and Nuclear Physics and the Office of Advanced Scientific
  Computing Research, of the U.S.  Department of Energy (DOE) under
  Contract Nos.  DE-FG02-92ER40704, DE-AC02-05CH11231,
  DE-FG02-06ER06-04, and DE-AC02-05CH11231; by a grant from the Gordon
  \& Betty Moore Foundation; by National Science Foundation Grant Nos.
  AST-0407297 (QUEST), and 0087344 \& 0426879 (HPWREN); by a Henri
  Chr\'etien International Research Grant administrated by the American
  Astronomical Society; the France-Berkeley Fund; by an Explora'Doc
  Grant by the R\'egion Rh\^one-Alpes.  We thank the AERONET principal
  investigator Brent Holben and his staff for its effort in establishing
  and maintaining the AERONET sites.  The Caltech Submillimeter
  Observatory is operated by the California Institute of Technology
  under cooperative agreement with the National Science Foundation
  (AST-0838261).
\end{acknowledgements}

%%%%%%%%%%%%%%%%
% Bibliography %
%%%%%%%%%%%%%%%%

\bibliographystyle{aa}
\bibliography{SNFextinction}

%%%%%%%%%%
% Tables %
%%%%%%%%%%

\clearpage

\begin{table*}
  \caption{List of the standard stars used in \snf for flux calibration
    purposes (\ie computation of the nightly atmospheric extinction, the
    nightly telluric correction template and the instrument
    calibration).}
  \label{tab:standard-stars}
  \centering
  \begin{tabular}{lcccccccc}
    \hline
    \hline
    Standard Star & Number of &R.A (J2000) &
    Dec. (J2000) & Magnitude ($V$) & Type\\
    &Observations&&&&\\
    \hline
    BD+174708  & 237 & 22 11 31.37 & +18 05 34.2 &  9.91 & sdF8  \\
    BD+254655  & 90  & 21 59 42.02 & +26 25 58.1 &  9.76 & O     \\
    BD+284211  & 92  & 21 51 11.07 & +28 51 51.8 & 10.51 & Op    \\
    BD+332642  & 66  & 15 51 59.86 & +32 56 54.8 & 10.81 & B2IV  \\
    BD+75325   & 75  & 08 10 49.31 & +74 57 57.5 &  9.54 & O5p   \\
    CD-32d9927 & 21  & 14 11 46.37 & -33 03 14.3 & 10.42 & A0    \\
    EG131      & 270 & 19 20 35.00 & -07 40 00.1 & 12.3  & DA    \\
    Feige110   & 145 & 23 19 58.39 & -05 09 55.8 & 11.82 & DOp   \\
    Feige34    & 128 & 10 39 36.71 & +43 06 10.1 & 11.18 & DO    \\
    Feige56    & 36  & 12 06 42.23 & +11 40 12.6 & 11.06 & B5p   \\
    Feige66    & 44  & 12 37 23.55 & +25 04 00.3 & 10.50 & sdO   \\
    Feige67    & 32  & 12 41 51.83 & +17 31 20.5 & 11.81 & sdO   \\
    G191B2B    & 102 & 05 05 30.62 & +52 49 54.0 & 11.78 & DA1   \\
    GD153      & 209 & 12 57 02.37 & +22 01 56.0 & 13.35 & DA1   \\
    GD71       & 165 & 05 52 27.51 & +15 53 16.6 & 13.03 & DA1   \\
    HD93521    & 227 & 10 48 23.51 & +37 34 12.8 &  7.04 & O9Vp  \\
    HR1544     & 226 & 04 50 36.69 & +08 54 00.7 &  4.36 & A1V   \\
    HR3454     & 159 & 08 43 13.46 & +03 23 55.1 &  4.30 & B3V   \\
    HR4468     & 127 & 11 36 40.91 & -09 48 08.2 &  4.70 & B9.5V \\
    HR4963     & 164 & 13 09 56.96 & -05 32 20.5 &  4.38 & A1IV  \\
    HR5501     & 202 & 14 45 30.25 & +00 43 02.7 &  5.68 & B9.5V \\
    HR718      & 253 & 02 28 09.54 & +08 27 36.2 &  4.28 & B9III \\
    HR7596     & 325 & 19 54 44.80 & +00 16 24.6 &  5.62 & A0III \\
    HR7950     & 236 & 20 47 40.55 & -09 29 44.7 &  3.78 & A1V   \\
    HR8634     & 255 & 22 41 27.64 & +10 49 53.2 &  3.40 & B8V   \\
    HR9087     & 210 & 00 01 49.42 & -03 01 39.0 &  5.12 & B7III \\
    HZ21       & 63  & 12 13 56.42 & +32 56 30.8 & 14.68 & DO2   \\
    HZ44       & 30  & 13 23 35.37 & +36 08 00.0 & 11.66 & sdO   \\
    LTT1020    & 45  & 01 54 49.68 & -27 28 29.7 & 11.52 & G     \\
    LTT1788    & 26  & 03 48 22.17 & -39 08 33.6 & 13.16 & F     \\
    LTT2415    & 57  & 05 56 24.30 & -27 51 28.8 & 12.21 & sdG   \\
    LTT377     & 41  & 00 41 46.82 & -33 39 08.2 & 11.23 & F     \\
    LTT3864    & 12  & 10 32 13.90 & -35 37 42.4 & 12.17 & F     \\
    LTT6248    & 32  & 15 39 00.02 & -28 35 33.1 & 11.80 & A     \\
    LTT9239    & 36  & 22 52 40.88 & -20 35 26.3 & 12.07 & F     \\
    LTT9491    & 56  & 23 19 34.98 & -17 05 29.8 & 14.11 & DC    \\
    NGC7293    & 39  & 22 29 38.46 & -20 50 13.3 & 13.51 & V.Hot \\
    P041C      & 31  & 14 51 58.19 & +71 43 17.3 & 12.00 & GV    \\
    P177D      & 119 & 15 59 13.59 & +47 36 41.8 & 13.47 & GV    \\
    \hline
  \end{tabular}
\end{table*}

\clearpage

\begin{table*}
  \caption{\snf median atmospheric extinction [mag/airmass], its
    variability (RMS and nMAD of all the individual extinctions) and its
    decomposition into physical components from 3200~\AA{} to 10000~\AA{}
    (100~\AA{} bins). A complete table with a 2~\AA{} binning is
    available in electronic form
    at the CDS via anonymous ftp to \url{cdsarc.u-strasbg.fr} (130.79.128.5)
    or via \url{http://cdsweb.u-strasbg.fr/cgi-bin/qcat?J/A+A/}.}
  \label{tab:atmospheric_extinction}
  \centering
  \begin{tabular}{ccccccc}
    \hline
    \hline
    Wavelength [\AA] & Total Extinction & \multicolumn{2}{c}{Variability} & Rayleigh & Ozone & Aerosols\\
    && RMS & nMAD &&&\\
    \hline
     3200 & 0.856 &   -   &   -   & 0.606 & 0.214 & 0.036\\
     3300 & 0.588 & 0.057 & 0.041 & 0.532 & 0.021 & 0.035\\
     3400 & 0.514 & 0.053 & 0.040 & 0.469 & 0.012 & 0.033\\
     3500 & 0.448 & 0.048 & 0.039 & 0.415 & 0.001 & 0.032\\
     3600 & 0.400 & 0.045 & 0.037 & 0.369 & 0.000 & 0.031\\
     3700 & 0.359 & 0.042 & 0.035 & 0.329 & 0.000 & 0.030\\
     3800 & 0.323 & 0.039 & 0.034 & 0.294 & 0.000 & 0.029\\
     3900 & 0.292 & 0.036 & 0.032 & 0.264 & 0.000 & 0.028\\
     4000 & 0.265 & 0.033 & 0.030 & 0.238 & 0.000 & 0.027\\
     4100 & 0.241 & 0.031 & 0.029 & 0.215 & 0.000 & 0.026\\
     4200 & 0.220 & 0.029 & 0.027 & 0.194 & 0.000 & 0.026\\
     4300 & 0.202 & 0.027 & 0.026 & 0.176 & 0.001 & 0.025\\
     4400 & 0.185 & 0.026 & 0.025 & 0.160 & 0.001 & 0.024\\
     4500 & 0.171 & 0.024 & 0.023 & 0.146 & 0.001 & 0.023\\
     4600 & 0.159 & 0.023 & 0.022 & 0.134 & 0.003 & 0.023\\
     4700 & 0.147 & 0.021 & 0.021 & 0.122 & 0.003 & 0.022\\
     4800 & 0.139 & 0.020 & 0.020 & 0.112 & 0.005 & 0.022\\
     4900 & 0.130 & 0.019 & 0.019 & 0.103 & 0.006 & 0.021\\
     5000 & 0.125 & 0.018 & 0.017 & 0.095 & 0.009 & 0.021\\
     5100 & 0.119 & 0.017 & 0.016 & 0.087 & 0.012 & 0.020\\
     5200 & 0.114 & 0.016 & 0.015 & 0.081 & 0.013 & 0.020\\
     5300 & 0.113 & 0.015 & 0.014 & 0.075 & 0.019 & 0.019\\
     5400 & 0.109 & 0.015 & 0.013 & 0.069 & 0.022 & 0.019\\
     5500 & 0.106 & 0.014 & 0.013 & 0.064 & 0.024 & 0.018\\
     5600 & 0.107 & 0.013 & 0.012 & 0.060 & 0.029 & 0.018\\
     5700 & 0.108 & 0.013 & 0.012 & 0.056 & 0.035 & 0.017\\
     5800 & 0.103 & 0.012 & 0.011 & 0.052 & 0.034 & 0.017\\
     5900 & 0.098 & 0.012 & 0.011 & 0.048 & 0.033 & 0.017\\
     6000 & 0.098 & 0.011 & 0.010 & 0.045 & 0.037 & 0.016\\
     6100 & 0.092 & 0.011 & 0.010 & 0.042 & 0.034 & 0.016\\
     6200 & 0.084 & 0.010 & 0.009 & 0.039 & 0.029 & 0.016\\
     6300 & 0.078 & 0.010 & 0.009 & 0.037 & 0.026 & 0.015\\
     6400 & 0.070 & 0.009 & 0.008 & 0.035 & 0.021 & 0.015\\
     6500 & 0.065 & 0.009 & 0.008 & 0.033 & 0.018 & 0.015\\
     6600 & 0.060 & 0.008 & 0.007 & 0.031 & 0.015 & 0.014\\
     6700 & 0.056 & 0.008 & 0.007 & 0.029 & 0.013 & 0.014\\
     6800 & 0.052 & 0.008 & 0.007 & 0.027 & 0.011 & 0.014\\
     6900 & 0.048 & 0.007 & 0.006 & 0.026 & 0.008 & 0.014\\
     7000 & 0.044 & 0.007 & 0.006 & 0.024 & 0.007 & 0.013\\
     7100 & 0.042 & 0.007 & 0.006 & 0.023 & 0.006 & 0.013\\
     7200 & 0.039 & 0.006 & 0.005 & 0.022 & 0.005 & 0.013\\
     7300 & 0.037 & 0.006 & 0.005 & 0.020 & 0.004 & 0.013\\
     7400 & 0.035 & 0.006 & 0.005 & 0.019 & 0.003 & 0.013\\
     7500 & 0.033 & 0.006 & 0.004 & 0.018 & 0.003 & 0.012\\
     7600 & 0.032 & 0.005 & 0.004 & 0.017 & 0.002 & 0.012\\
     7700 & 0.030 & 0.005 & 0.004 & 0.016 & 0.002 & 0.012\\
     7800 & 0.029 & 0.005 & 0.004 & 0.016 & 0.002 & 0.012\\
     7900 & 0.028 & 0.005 & 0.004 & 0.015 & 0.002 & 0.012\\
     8000 & 0.027 & 0.005 & 0.003 & 0.014 & 0.001 & 0.011\\
     8100 & 0.026 & 0.004 & 0.003 & 0.013 & 0.001 & 0.011\\
     8200 & 0.025 & 0.004 & 0.003 & 0.013 & 0.001 & 0.011\\
     8300 & 0.024 & 0.004 & 0.003 & 0.012 & 0.001 & 0.011\\
     8400 & 0.023 & 0.004 & 0.003 & 0.012 & 0.001 & 0.011\\
     8500 & 0.023 & 0.004 & 0.002 & 0.011 & 0.001 & 0.011\\
     8600 & 0.022 & 0.004 & 0.002 & 0.011 & 0.001 & 0.010\\
     8700 & 0.021 & 0.004 & 0.002 & 0.010 & 0.001 & 0.010\\
     8800 & 0.021 & 0.004 & 0.002 & 0.010 & 0.001 & 0.010\\
     8900 & 0.020 & 0.003 & 0.002 & 0.009 & 0.001 & 0.010\\
     9000 & 0.019 & 0.003 & 0.002 & 0.009 & 0.001 & 0.010\\
     9100 & 0.019 & 0.003 & 0.002 & 0.008 & 0.001 & 0.010\\
     9200 & 0.018 & 0.003 & 0.002 & 0.008 & 0.001 & 0.010\\
     9300 & 0.018 & 0.003 & 0.002 & 0.008 & 0.001 & 0.009\\
     9400 & 0.017 & 0.003 & 0.001 & 0.007 & 0.001 & 0.009\\
     9500 & 0.017 & 0.003 & 0.001 & 0.007 & 0.000 & 0.009\\
     9600 & 0.016 & 0.003 & 0.001 & 0.007 & 0.000 & 0.009\\
     9700 & 0.016 & 0.003 & 0.001 & 0.006 & 0.000 & 0.009\\
     9800 & 0.015 &   -   &   -   & 0.006 & 0.000 & 0.009\\
     9900 & 0.015 &   -   &   -   & 0.006 & 0.000 & 0.009\\
    10000 & 0.014 &   -   &   -   & 0.006 & 0.000 & 0.009\\
    \hline
  \end{tabular}
\end{table*}

\clearpage

\end{document}